\documentclass[aip,jmp,preprint,amsmath,amssymb,eqsecnum,showpacs]{revtex4-1}

\usepackage{graphicx}

\newtheorem{theorem}{Theorem}

\newcommand{\ov}[1]{{\bar{#1}}}
\newcommand{\bmath}[1]{\mbox{{\boldmath{{$#1$}}}}}
\newcommand{\drs}{\partial _{r_{*}}}
\newcommand{\Diag}{\mathrm {Diag}}
\newcommand{\Tr}{\mathrm {Tr}}

\begin{document}

\title{On the stability of soliton and hairy black hole
solutions of ${\mathfrak {su}}(N)$ Einstein-Yang-Mills theory
with a negative cosmological constant}

\author{J. Erik Baxter}
\email{E.Baxter@shu.ac.uk}
\affiliation{
Norfolk Building,
Sheffield Hallam University,
1 Howard Street, Sheffield. S1 1WB United Kingdom
}

\author{Elizabeth Winstanley}
\email{E.Winstanley@sheffield.ac.uk}
\affiliation{
Consortium for Fundamental Physics,
School of Mathematics and Statistics, \\
The University of Sheffield,
Hicks Building,
Hounsfield Road,
Sheffield.
S3 7RH
United Kingdom}

\begin{abstract}
We investigate the stability of spherically symmetric, purely magnetic,
soliton and black hole solutions of four-dimensional
${\mathfrak {su}}(N)$ Einstein-Yang-Mills theory with a negative cosmological constant $\Lambda $.
These solutions are described by $N-1$ magnetic gauge field functions $\omega _{j}$.
We consider linear, spherically symmetric, perturbations of these solutions.
The perturbations decouple into two sectors, known as the sphaleronic and gravitational sectors.
For any $N$, there are no instabilities in the sphaleronic sector if all the magnetic
gauge field functions $\omega _{j}$ have no zeros, and satisfy a set of $N-1$ inequalities.
In the gravitational sector, we prove that there are solutions which have no instabilities in a neighbourhood of stable embedded ${\mathfrak {su}}(2)$ solutions, provided the magnitude of the cosmological constant
$\left| \Lambda \right| $ is sufficiently large.
\end{abstract}


\maketitle

\section{Introduction}
\label{sec:intro}

Soliton and black hole solutions of Einstein-Yang-Mills (EYM) theory have been studied extensively for over twenty years (see, for example, Ref.~\onlinecite{Volkov1} for a review).
The first solutions found were spherically symmetric, purely magnetic, asymptotically flat, solitons \cite{Bartnik1} and black holes \cite{Bizon1} in four-dimensional
${\mathfrak {su}}(2)$ EYM.
Discrete families of solutions were found numerically and their existence was later proven (see Refs.~\onlinecite{Breitenlohner1, Smoller} for some analytic work).
The purely magnetic gauge field is described by a single function $\omega $, which has at least one zero.  The families of solutions are characterized by the event horizon radius $r_{h}$ (with $r_{h}=0$ corresponding to soliton solutions) and the number of zeros of the function $\omega $.
Both the soliton and black hole families of solutions are unstable under linear, spherically symmetric perturbations \cite{Straumann1},
with the number of unstable perturbation modes of the solutions being twice the number of zeros of
of $\omega $ \cite{Lavrelashvili1,Volkov2}.

Many generalizations of the original spherically symmetric ${\mathfrak {su}}(2)$ solitons and black holes have been considered in the literature (some of which are reviewed in Ref.~\onlinecite{Volkov1}).
For example, numerical solutions have been found which retain the spherical symmetry of the original solutions but enlarge the gauge group to ${\mathfrak {su}}(N)$ (see, for example, Ref.~\onlinecite{Galtsov1}).
The solution space is more complicated with the larger gauge group, but solutions still exist in discrete families.
Furthermore, all asymptotically flat, spherically symmetric, soliton and black hole solutions with arbitrary gauge group are unstable under linear, spherically symmetric perturbations \cite{Brodbeck}.

The model can also be generalized by considering space-times which are not asymptotically flat or which have more than four dimensions.  In four-dimensional space-time, discrete families of spherically symmetric soliton and black hole solutions of ${\mathfrak {su}}(2)$ EYM also exist in asymptotically de Sitter space-time \cite{Volkov3}, but, like their asymptotically flat counterparts, they are unstable \cite{dSstab}.
If one considers higher-dimensional space-times, in order to have spherically symmetric finite mass solutions, the YM action must be modified by the addition
of higher-order curvature terms \cite{Volkov4}.
With these additional terms, soliton and black hole solutions have been found in both asymptotically flat and asymptotically de Sitter space-times \cite{HD}.

About ten years after the discovery of four-dimensional, spherically symmetric, purely magnetic, asymptotically flat, solitons and black holes in ${\mathfrak {su}}(2)$ EYM, their analogues in four-dimensional asymptotically anti-de Sitter (adS) space-time were found \cite{Winstanley1,Bjoraker,Breitenlohner2}.
The purely magnetic ${\mathfrak {su}}(2)$ gauge field is still described by a single function $\omega $, but now continuous families of solutions are found,  which are indexed by the event horizon radius $r_{h}$ as before (including $r_{h}=0$ for solitons), the negative cosmological constant $\Lambda $ and the value of the gauge field function on the horizon $\omega _{h}$ (there is an alternative parameter for soliton solutions, which governs the behaviour of the magnetic gauge field function near the origin).
One striking feature of the families of solutions is the existence, for sufficiently large $\left| \Lambda \right| $, of solutions where the magnetic gauge field function $\omega $ has no zeros.
These solutions where $\omega $ is nodeless are particularly important because at least some of them are stable under linear, spherically symmetric, perturbations \cite{Winstanley1,Bjoraker,Breitenlohner2}.
The existence, for sufficiently large $\left| \Lambda \right| $, of soliton and black hole solutions which are stable under linear, non-spherically symmetric, perturbations has also been proven \cite{Sarbach1, Sarbach2}.
In this paper we consider only four-dimensional, spherically symmetric solutions, but asymptotically adS generalizations to higher-dimensions \cite{HDadS}
or non-spherically symmetric space-times \cite{adSgen,adSthermo} do exist.

One generalization which has received a great deal of attention in the literature over the past seven years is topological EYM black holes in adS, in particular the relevance of black holes with planar event horizons to models of holographic superconductors (see, for example, the recent review \cite{holreview} for more details and references).
Purely magnetic black holes with non-spherical event horizon topology in ${\mathfrak {su}}(2)$ EYM in adS appeared in the literature soon after their spherically symmetric counterparts \cite{adStop}.
Unlike the situation in asymptotically flat space-time \cite{ershov}, in asymptotically adS space-time ${\mathfrak {su}}(2)$ EYM black holes and solitons can
have nontrivial electric and magnetic fields.
While spherically symmetric dyonic solutions (both solitons and black holes) in ${\mathfrak {su}}(2)$ EYM in adS were found soon after the purely magnetic  black holes \cite{Bjoraker}, topological dyonic solutions have been studied only more recently.
Gubser \cite{gubser} considered four-dimensional dyonic ${\mathfrak {su}}(2)$ EYM black holes in adS with planar event horizons.
He found a second-order phase transition between the embedded planar Reissner-Nordstr\"om-adS black hole and a black hole with a nontrivial
YM field condensate.
Planar EYM black holes in adS have subsequently been widely studied as models of $p$-wave superconductors \cite{pufu} (see also Refs.~\onlinecite{holreview,holographic}
for a selection of work in this area).

Returning to four-dimensional, spherically symmetric, purely magnetic, asymptotically adS solutions, a natural question is whether the above stable ${\mathfrak {su}}(2)$ solitons and black holes have generalizations with a larger ${\mathfrak {su}}(N)$ gauge group.
The answer is affirmative: such solutions have been found numerically for gauge groups ${\mathfrak {su}}(3)$ and ${\mathfrak {su}}(4)$ \cite{Baxter2}.
For the larger ${\mathfrak {su}}(N)$ gauge group, the purely magnetic gauge field is described by $N-1$ functions $\omega _{j}$ (see section \ref{sec:ansatz} below).
As in the ${\mathfrak {su}}(2)$ case, there are continuous families of solutions, parameterized by the negative cosmological constant $\Lambda $, the event horizon radius $r_{h}$  (with $r_{h}=0$ for soliton solutions) and $N-1$ parameters describing the form of the gauge field functions either on the event horizon or near the origin.
Numerically it is found that, if $\left| \Lambda \right| $ is sufficiently large, then there are solutions in which all the gauge field functions $\omega _{j}$ have no zeros.
For general $N$, the existence of such nodeless, spherically symmetric, purely magnetic, asymptotically adS, soliton and black hole solutions of ${\mathfrak {su}}(N)$ EYM has been proven for
sufficiently large $\left| \Lambda \right| $ \cite{Baxter3}.

In this paper we address the question of whether these soliton and black hole solutions of ${\mathfrak {su}}(N)$ EYM in which all the magnetic gauge field functions have no zeros are stable.
The outline of the paper is as follows.
In section \ref{sec:EYM} we introduce ${\mathfrak {su}}(N)$ EYM with a negative cosmological constant and the ansatz \cite{Kunzle1} for a spherically symmetric gauge potential. We derive the field equations describing static, purely magnetic, configurations and the perturbation equations for linear, spherically symmetric, perturbations.
With an appropriate choice of gauge, the perturbation equations decouple into two sectors: the sphaleronic and gravitational sectors.
These are considered in sections \ref{sec:sphaleronic} and \ref{sec:gravitational}
respectively.
Finally we present our conclusions in section \ref{sec:conc}.

\section{The Einstein-Yang-Mills equations}
\label{sec:EYM}

\subsection{Action, metric and gauge potential}
\label{sec:ansatz}

The action for four-dimensional ${\mathfrak {su}}(N)$ Einstein-Yang-Mills (EYM) theory with a negative cosmological constant $\Lambda <0$ is:
\begin{equation}
S_{\mathrm {EYM}} = \frac {1}{2} \int d ^{4}x {\sqrt {-g}} \left[
R - 2\Lambda - \Tr \, F_{\mu \nu }F ^{\mu \nu }
\right] ,
\label{eq:action}
\end{equation}
where $R$ is the Ricci scalar, $F_{\mu \nu }$ is the non-Abelian gauge field
and ${\mathrm {Tr}}$ denotes a Lie algebra trace.
Throughout this paper, the metric has signature $\left( -, +, +, + \right) $ and
we use units in which $4\pi G = 1 = c$.
In addition, the gauge coupling constant is fixed to be equal to unity.
Varying the action (\ref{eq:action}) yields the field equations:
\begin{eqnarray}
T_{\mu \nu } & = & R_{\mu \nu } - \frac {1}{2} R g_{\mu \nu } +
\Lambda g_{\mu \nu },
\nonumber \\
0 & = & D_{\mu } F_{\nu }{}^{\mu } = \nabla _{\mu } F_{\nu }{}^{\mu }
+ \left[ A_{\mu }, F_{\nu }{}^{\mu } \right] ;
\label{eq:fieldeqns}
\end{eqnarray}
where the Yang-Mills stress-energy tensor is
\begin{equation}
T_{\mu \nu } = \Tr \left[ F_{\mu \lambda } F_{\nu }{}^{\lambda }
- \frac {1}{4} g_{\mu \nu }
 F_{\lambda \sigma} F^{\lambda \sigma } \right] ,
\label{eq:Tmunu}
\end{equation}
which involves a Lie-algebra trace.
The Yang-Mills gauge field $F_{\mu \nu }$ is given in terms of the gauge potential
$A_{\mu }$ by
\begin{equation}
F_{\mu \nu } = \partial _{\mu }A_{\nu } - \partial _{\nu }A_{\mu } +
\left[ A_{\mu },A_{\nu } \right] .
\end{equation}

Our focus in this paper is on equilibrium, static, spherically symmetric, soliton and black hole solutions of the field equations (\ref{eq:fieldeqns}) and time-dependent, spherically symmetric, perturbations of those equilibrium solutions.
We therefore consider a time-dependent, spherically symmetric, geometry, whose metric in standard Schwarzschild-like co-ordinates takes the form
\begin{equation}
ds^{2} = - \mu S^{2} \, dt^{2} + \mu ^{-1} \, dr^{2} +
r^{2} \, d\theta ^{2} + r^{2} \sin ^{2} \theta \, d\phi ^{2} ,
\label{eq:metric}
\end{equation}
where the metric functions $\mu (t,r)$ and $S(t,r)$ depend on the co-ordinates $t$ and $r$ only.
Since we have a negative cosmological constant $\Lambda $, it is useful to write the metric function $\mu (t,r)$ in the form
\begin{equation}
\mu (t,r) = 1 - \frac {2m(t,r)}{r} - \frac {\Lambda r^{2}}{3}.
\label{eq:mu}
\end{equation}
In our later analysis we will also find it useful to define another function
$\Delta (t,r)$ such that
\begin{equation}
S(t,r) = \exp \Delta (t,r).
\end{equation}
With this metric ansatz the relevant components of the Einstein tensor are:
\begin{eqnarray}
G_{tt} & = &
-\frac{\mu S^2}{r^2}\left(\mu'r-1+\mu\right) ,
\nonumber
\\
G_{tr} & = & -\frac{\dot{\mu}}{\mu r} ,
\nonumber
\\
G_{rr} & = & \frac{1}{\mu Sr^2}\left(\mu'Sr+2S'\mu r-S+\mu S\right) ,
\label{eq:EinsteinTensor}
\end{eqnarray}
where here and throughout this paper we use a dot to denote $\partial /\partial t$
and a prime to denote $\partial /\partial r$.
Note that we do not need to consider the $G_{\theta \theta }$ or $G_{\phi \phi }$
components of the Einstein tensor as the field equations involving these components follow from those involving the components in (\ref{eq:EinsteinTensor}) by the Bianchi identities.

We make the following ansatz for a time-dependent, spherically symmetric,
${\mathfrak {su}}(N)$ gauge potential \cite{Kunzle1}:
\begin{eqnarray}
A  & = &
{\mathcal {A}} \, dt + {\mathcal {B}} \, dr +
\frac {1}{2} \left( C - C^{H} \right) \, d\theta
- \frac {i}{2} \left[ \left(
C + C^{H} \right) \sin \theta + D \cos \theta \right] \, d\phi ,
\label{eq:gaugepot}
\end{eqnarray}
where ${\mathcal {A}}$, ${\mathcal {B}}$, $C$ and $D$ are all $\left( N \times N \right) $ matrices depending on the co-ordinates $(t,r)$ only and
$C^{H}$ is the Hermitian conjugate of $C$.
With this gauge potential ansatz, the non-zero components of the gauge field are:
\begin{eqnarray}
F_{tr} & = & \dot{B}-A' ,
\nonumber
\\
F_{t\theta}  & = & \frac{1}{2}\left\{ (C-C^H)\dot{}+[A,C-C^H]\right\} ,
\nonumber
\\
F_{t\phi} & = & -\frac{i}{2}\left\{(C+C^H)\dot{}+[A,C+C^H]\right\}\sin\theta ,
\nonumber
\\
F_{r\theta} & = & \frac{1}{2}\left\{(C-C^H)'+[B,C-C^H]\right\} ,
\nonumber
\\
F_{r\phi} & = & -\frac{i}{2}\left\{(C+C^H)'+[B,C+C^H]\right\}\sin\theta ,
\nonumber
\\
F_{\theta\phi} & = & -\frac{i}{2}\left\{ [C,C^H]-D\right\}\sin\theta.
\label{eq:Fmunu}
\end{eqnarray}
In computing the component $F_{\theta \phi }$ we have made use of the identities
\cite{Kunzle1}
\begin{equation}
[D,C] =2C ,
\qquad
[D, C^H]=-2C^H.
\label{eq:Dcomm}
\end{equation}
The matrices ${\mathcal {A}}$ and ${\mathcal {B}}$ are diagonal and traceless, and we define functions $\alpha _{j}(t,r)$ and $\beta _{j}(t,r)$ for $j=1,\ldots N$ such that
\begin{eqnarray}
{\mathcal {A}} & = & i \, \Diag   \left( \alpha _{1}(t,r), \ldots
\alpha _{N}(t,r) \right) ,
\nonumber \\
{\mathcal {B}} & = & i \, \Diag   \left( \beta _{1}(t,r), \ldots
\beta _{N}(t,r) \right),
\end{eqnarray}
where the fact that these two matrices must be traceless means that
\begin{equation}
\sum _{j=1}^{N} \alpha _{j} (t,r) = 0 = \sum _{j=1}^{N} \beta _{j}(t,r).
\end{equation}
The matrix $C$ is upper triangular, with non-zero entries only immediately above the main diagonal. These entries are given in terms of functions $\omega _{j}(t,r)$
and $\gamma _{j}(t,r)$ for $j=1,\ldots , N-1$ by
\begin{equation}
C_{j,j+1}=\omega _{j}(t,r) e^{i\gamma _{j}(t,r)}.
\label{eq:Celements}
\end{equation}
Finally, the matrix $D$ is a constant diagonal matrix \cite{Kunzle1}:
\begin{equation}
D=\Diag \left(N-1,N-3,\ldots,-N+3,-N+1\right) .
\label{eq:matrixD}
\end{equation}

\subsection{Static solutions}
\label{sec:static}

For static solutions, all field variables depend only on the radial co-ordinate $r$.
We denote static equilibrium functions with a bar (e.g.~${\ov {\omega }}_{j}$) to distinguish them from the time-dependent perturbations which we shall consider shortly.
The static equilibrium solutions in which we are interested are purely magnetic, which means that we set the electric gauge field functions ${\ov {\alpha }}_{j}(r)\equiv 0$ for all $j=1, \ldots ,N$.
The remaining gauge freedom is then used to set all the functions
${\ov {\beta }}_{j}(r)\equiv 0$ for $j=1,\ldots ,N$ \cite{Kunzle1}.
From now on we assume that none of the magnetic gauge functions
${\ov {\omega }}_{j}(r)$ are identically zero.
In asymptotically flat space, other families of solutions have been found when this assumption is relaxed \cite{Galtsov1}.
Assuming that none of the ${\ov {\omega }}_{j}(r)$ are identically zero, one of the Yang-Mills equations becomes \cite{Kunzle1}:
\begin{equation}
{\ov {\gamma }}_{j}(r) =0, \qquad j=1,\ldots , N-1 ,
\end{equation}
and the gauge field is described by the $N-1$ magnetic gauge field functions
${\ov {\omega }}_{j}(r)$, $j=1,\ldots , N-1$.
We comment that our ansatz (\ref{eq:gaugepot})
is by no means the only possible choice in
${\mathfrak {su}}(N)$ EYM (in Ref.~\onlinecite{Bartnik2} all irreducible models are explicitly listed for $N\le 6$, and
techniques for finding {\em {all}} spherically symmetric ${\mathfrak {su}}(N)$ gauge potentials are developed).

\subsubsection{Static field equations}
\label{sec:staticeqns}

For purely magnetic, static equilibrium solutions as described above, the field equations (\ref{eq:fieldeqns}) simplify as follows.
The Einstein equations take the form:
\begin{equation}
{\ov {m}} ' = {\ov {\mu }} (r) {\ov {\Gamma }} + r^{2} {\ov {\Pi }},
\qquad
{\ov {\Delta }} ' = \frac {2{\ov {\Gamma }}}{r} ,
\label{eq:Ee}
\end{equation}
where
\begin{eqnarray}
{\ov {\Gamma }} & = & \sum _{j=1}^{N-1} {\ov {\omega }}_{j}'^{2},
\nonumber \\
{\ov {\Pi }} & = & \frac {1}{4r^{4}} \sum _{j=1}^{N} \left[ \left(
{\ov {\omega }}_{j}^{2} - {\ov {\omega }}_{j-1}^{2}
-N - 1+ 2j\right) ^{2} \right] .
\label{eq:Pistatic}
\end{eqnarray}
The $N-1$ Yang-Mills equations take the form
\begin{equation}
0 = r^{2} {\ov {\mu }} {\ov {\omega }}_{j}'' + \left(
2{\ov {m}} - 2r^{3} {\ov {\Pi}} - \frac {2\Lambda r^{3}}{3} \right)
{\ov {\omega }}_{j} + W_{j} {\ov {\omega }}_{j} ,
\label{eq:YMe}
\end{equation}
where
\begin{equation}
W_{j} = 1-{\ov {\omega }}_{j}^{2} + \frac {1}{2} \left(
{\ov {\omega }}_{j-1}^{2} + {\ov {\omega }}_{j+1}^{2} \right) .
\label{eq:Wdef}
\end{equation}
The field equations (\ref{eq:Ee}, \ref{eq:YMe}) are invariant under the transformation
\begin{equation}
{\ov {\omega }}_{j} (r) \rightarrow -{\ov {\omega }}_{j}(r)
\label{eq:omegatransform}
\end{equation}
independently for each $j$, and also under the substitution:
\begin{equation}
j\rightarrow N-j.
\end{equation}

\subsubsection{Boundary conditions}
\label{sec:boundary}

The field equations (\ref{eq:Ee}, \ref{eq:YMe}) are singular at the origin $r=0$,
a black hole event horizon $r=r_{h}$ (where ${\ov {\mu }}(r_{h})=0$) and at
infinity $r\rightarrow \infty $.
Below we briefly outline the form of the equilibrium field functions in a neighbourhood of the singular points.
The existence of local solutions near these singular points, with the forms below, is proven in Ref.~\onlinecite{Baxter3}.

\paragraph{Origin}

The form of the static field functions near the origin is rather complicated.
In particular, to completely specify the gauge field in a neighbourhood of the origin, a power series up to $O(r^{N})$ is required,
involving $N-1$ initial parameters. These $N-1$ parameters, together with the cosmological constant $\Lambda $, completely determine the solution in a neighbourhood of the origin.
The details of this power series can be found in Ref.~\onlinecite{Baxter3}
(following the analysis of Ref.~\onlinecite{Kunzle2} for the asymptotically flat case).
For our analysis in this paper, we only require the leading order behaviour
of the static field functions, which is:
\begin{eqnarray}
{\ov {m}}(r) & = &  m_{3}r^{3} + O(r^{4}),
\nonumber \\
{\ov {S}}(r) & = &  S_{0} + S_{2}r^{2} + O(r^{3}),
\nonumber \\
{\ov {\omega }}_{j}(r) &  = &  \pm \left[ j \left( N - j\right) \right] ^{\frac {1}{2}}
+ O(r^{2}),
\label{eq:origin1}
\end{eqnarray}
where $m_{3}$, $S_{0}$ and $S_{2}$ are constants.
Without loss of generality, we take the positive sign in the form of
${\ov {\omega }}_{j}(r)$ due to the invariance of the field equations under the
transformation (\ref{eq:omegatransform}).

\paragraph{Event horizon}

Assuming there is a non-extremal black hole event horizon at $r=r_{h}$, the metric
function ${\ov {\mu }}(r)$ will have a single zero there.
This fixes the value of ${\ov {m}}(r_{h})$ to be
\begin{equation}
{\ov {m}}( r_{h} ) = \frac {r_{h}}{2} - \frac {\Lambda r_{h}^{3}}{6}.
\end{equation}
In a neighbourhood of the horizon, the field variables have the form
\begin{eqnarray}
{\ov {m}}(r)  & = &  {\ov {m}} (r_{h}) + {\ov {m}}' (r_{h}) \left( r - r_{h} \right)
+ O \left( r- r_{h} \right) ^{2} ,
\nonumber \\
{\ov {\omega }}_{j} (r) &  = &  {\ov {\omega }}_{j}(r_{h})
+ {\ov {\omega }}_{j}' (r_{h})
\left( r - r_{h} \right)
+ O \left( r -r_{h} \right) ^{2},
\nonumber \\
{\ov {S}}(r)  & = &  {\ov {S}}(r_{h}) + {\ov {S}}'(r_{h}) \left( r-r_{h} \right)
+ O\left( r - r_{h} \right) ,
\label{eq:horizon}
\end{eqnarray}
where ${\ov {m}}'(r_{h})$, ${\ov {\omega }}_{j}'(r_{h})$ and ${\ov {S}}'(r_{h})$ can be written
in terms of the constants ${\ov {\omega }}_{j}(r_{h})$ and $S(r_{h})$ by using the field equations (\ref{eq:Ee}, \ref{eq:YMe}).
Again, due to the invariance of the field equations under the transformation
(\ref{eq:omegatransform}), we may take ${\ov {\omega }}_{j}(r_{h})>0$ without loss of generality.
The $N-1$ initial parameters $\omega _{j}(r_{h})$, together with the cosmological constant $\Lambda $ and event horizon radius $r_{h}$, completely
determine the solution of the field equations in a neighbourhood of the horizon \cite{Baxter3}.

\paragraph{Infinity}

As $r\rightarrow \infty $, the field variables have the form:
\begin{eqnarray}
{\ov {m}}(r) &   =  &  M + O \left( r^{-1} \right) ,
\nonumber \\
{\ov {S}}(r) & = & 1 + O\left( r^{-1} \right) ,
\nonumber \\
{\ov {\omega }}_{j}(r) &  =  & {\ov {\omega }}_{j,\infty } + c_{j}r^{-1} +
O \left( r^{-2} \right) ,
\label{eq:infinity}
\end{eqnarray}
where $M$, ${\ov {\omega }}_{j,\infty }$ and $c_{j}$ are constants.

\subsubsection{Embedded solutions}
\label{sec:embedded}

Despite the complexity of the static field equations (\ref{eq:Ee}, \ref{eq:YMe}),
there are some embedded solutions which will be useful in our later analysis.

\paragraph{Schwarzschild-adS}

If we set
\begin{equation}
{\ov {\omega }}_{j} (r) \equiv \pm {\sqrt {j\left( N-j \right) }},
\end{equation}
for all $j=1,\ldots , N-1$ then the components of the gauge field strength tensor (\ref{eq:Fmunu}) vanish identically.
In this case the stress-energy tensor (\ref{eq:Tmunu}) therefore also vanishes and
we obtain the Schwarzschild-adS black hole solution
with
\begin{equation}
{\ov {m}}(r) \equiv M, \qquad
{\ov {S}}(r) \equiv 1,
\end{equation}
where $M$ is a constant representing the mass of the black hole.
Setting $M=0$ gives pure adS space-time as a solution of the field equations.

\paragraph{Reissner-Nordstr\"om-adS}

Alternatively, if we set
\begin{equation}
{\ov {\omega }}_{j}(r) \equiv 0
\end{equation}
for all $j=1,\ldots, N-1$ then the gauge field strength tensor (\ref{eq:Fmunu}) does not vanish as $F_{\theta \phi }$ has a contribution from the nonzero matrix
$D$ (\ref{eq:matrixD}).
In this case we obtain the magnetically charged Reissner-Nordstr\"om black hole solution with
\begin{equation}
{\ov {m}}(r) = M - \frac {Q^{2}}{2r}, \qquad
{\ov {S}}(r) \equiv 1,
\end{equation}
where the magnetic charge $Q$ is fixed to be
\begin{equation}
Q^{2} = \frac {1}{6} N \left( N+1 \right) \left( N-1 \right) .
\end{equation}

\paragraph{Embedded ${\mathfrak {su}}(2)$ solutions}

The above two solutions are effectively Abelian embedded solutions.
For all $N>2$, there is another class of embedded solutions, corresponding to
${\mathfrak {su}}(2)$ non-Abelian solutions.
To obtain these solutions, we write the $N-1$ magnetic gauge field functions
${\ov {\omega }}_{j}(r)$ in terms of a single magnetic gauge field function
${\ov {\omega }}(r)$ as follows:
\begin{equation}
{\ov {\omega }}_{j}(r) = \pm {\ov {\omega }}(r)  {\sqrt {j\left( N - j\right)}} .
\end{equation}
It is shown in Ref.~\onlinecite{Baxter3} that, by a suitable rescaling of the other field variables, in this case the static field equations (\ref{eq:Ee}, \ref{eq:YMe}) reduce to those for the ${\mathfrak {su}}(2)$ case with ${\ov {\omega }}(r)$ as the single magnetic gauge field function.
Therefore any solution of the ${\mathfrak {su}}(2)$ field equations can be embedded as a solution of the ${\mathfrak {su}}(N)$ field equations.
In particular, setting ${\bar {\omega }}(r) \equiv 1$ gives the Schwarzschild-adS solution of the embedded ${\mathfrak {su}}(2)$ field equations and setting
${\bar {\omega }}(r) \equiv 0$ gives the magnetically charged Reissner-Nordstr\"om-adS black hole.

\subsubsection{Non-embedded solutions}
\label{sec:numerical}

Genuinely ${\mathfrak {su}}(N)$ static soliton and black hole solutions, which do not fall into one of the categories described in section \ref{sec:embedded}, have been studied in some detail already in the literature \cite{Baxter2}.  Therefore in this section we very briefly describe some of the key features of the solutions which are required for our subsequent analysis.

As discussed in the introduction, continuous families of solutions of the field equations (\ref{eq:Ee}, \ref{eq:YMe}) are found numerically. The solutions are parameterized by the cosmological constant $\Lambda $, the event horizon radius $r_{h}$ (we can consider $r_{h}=0$ to represent soliton solutions) and, for
${\mathfrak {su}}(N)$, there are a further $N-1$ parameters which describe the gauge field (see section \ref{sec:boundary}).
For black holes, these $N-1$ parameters are simply the values of the gauge field functions on the horizon $\omega _{j}(r_{h})$ (\ref{eq:horizon}).
For soliton solutions the situation is more complicated, details of the parameters in this case can be found in Refs.~\onlinecite{Baxter2,Baxter3}.

\begin{figure}
\includegraphics[width=7.5cm]{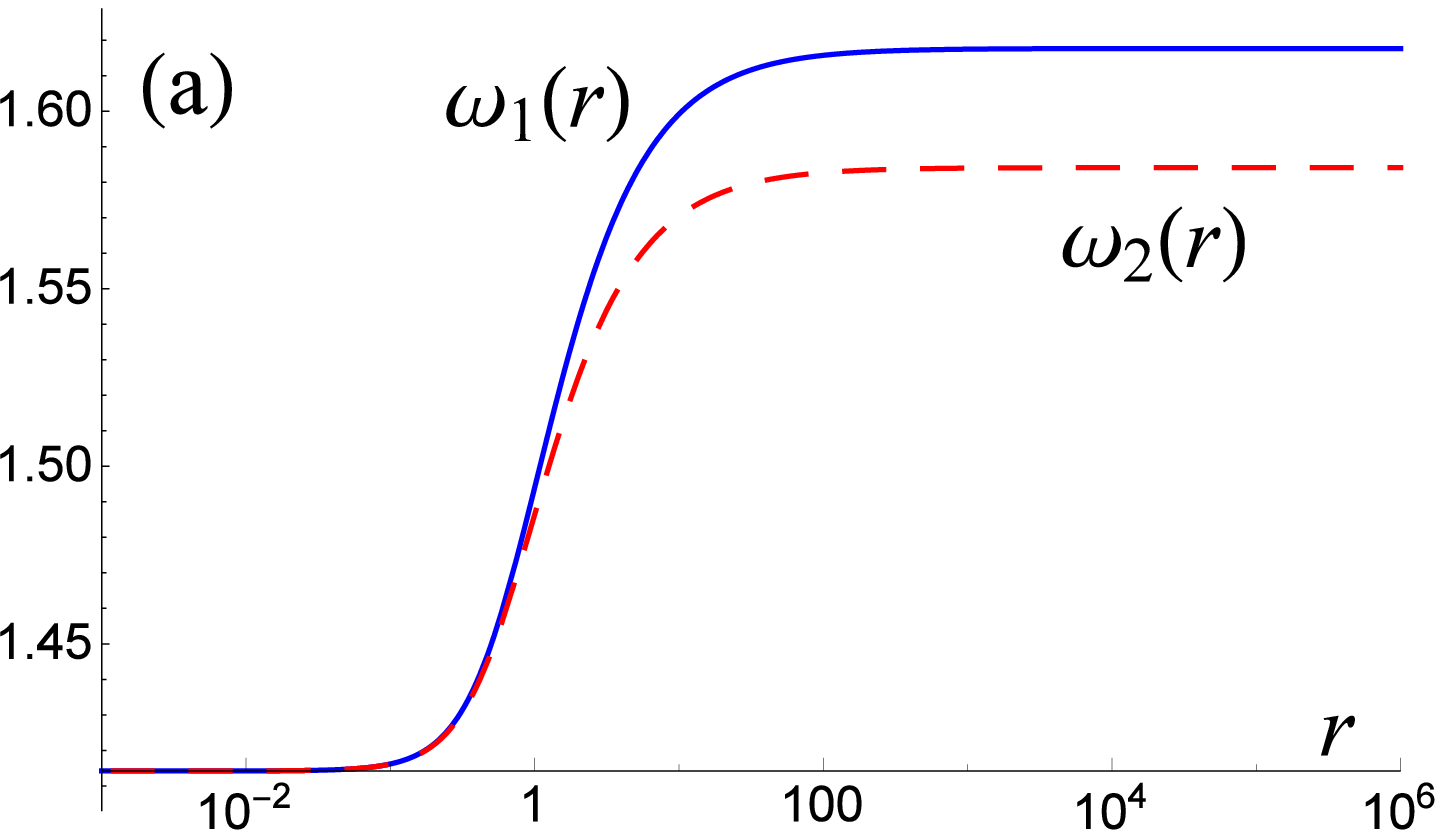}
\includegraphics[width=7.5cm]{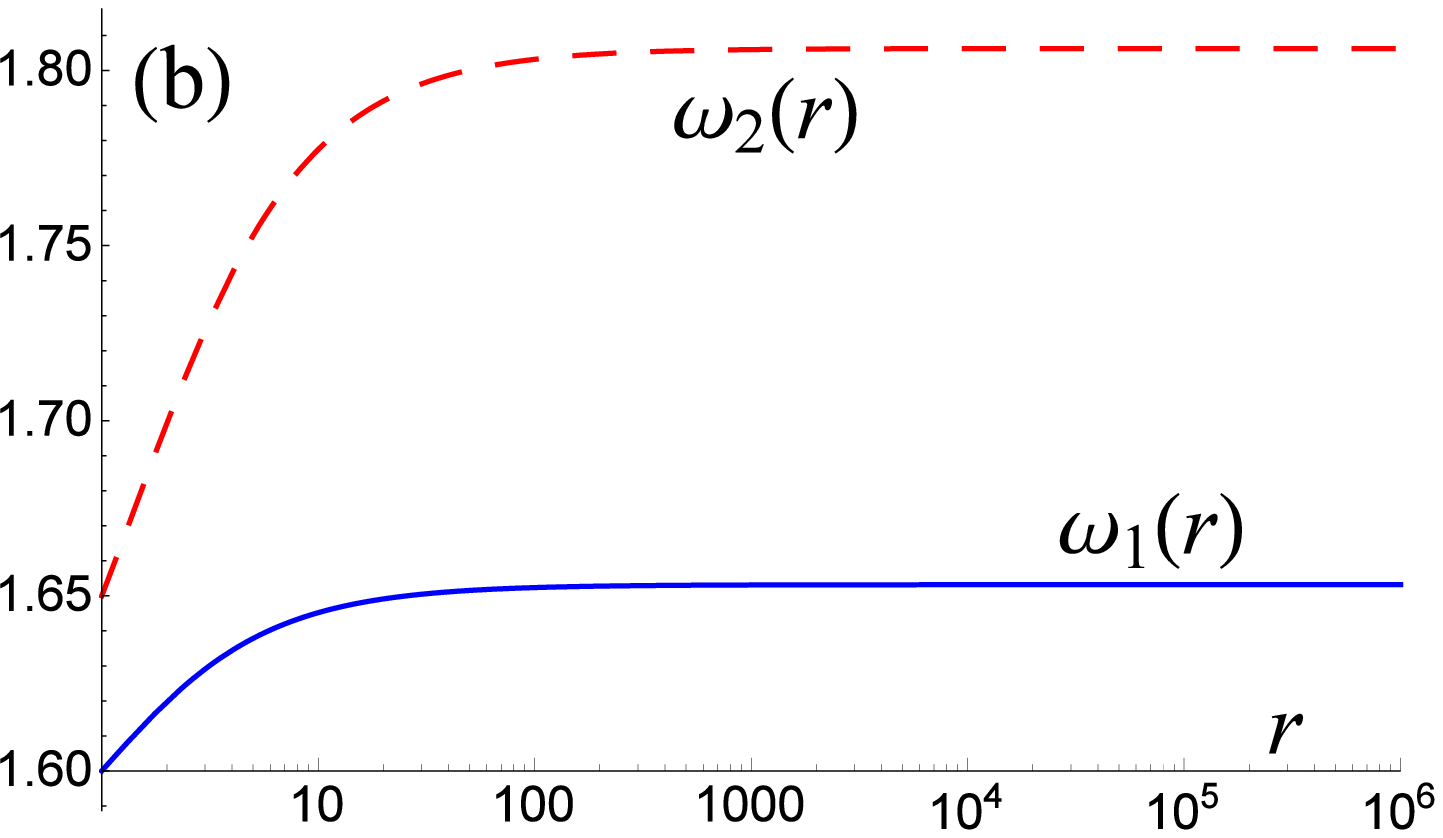}
\caption{Example nodeless solutions for ${\mathfrak {su}}(3)$ EYM with $\Lambda = -10$.  In each case we plot the gauge field functions $\omega _{1}(r)$
and $\omega _{2}(r)$ (the typical behaviour of the metric functions can be found in the example solutions plotted in Ref.~\onlinecite{Baxter2}).
In (a) we show a soliton solution, and in (b) a black hole solution with $r_{h}=1$.}
\label{fig:1}
\end{figure}

\begin{figure}
\includegraphics[width=7.5cm]{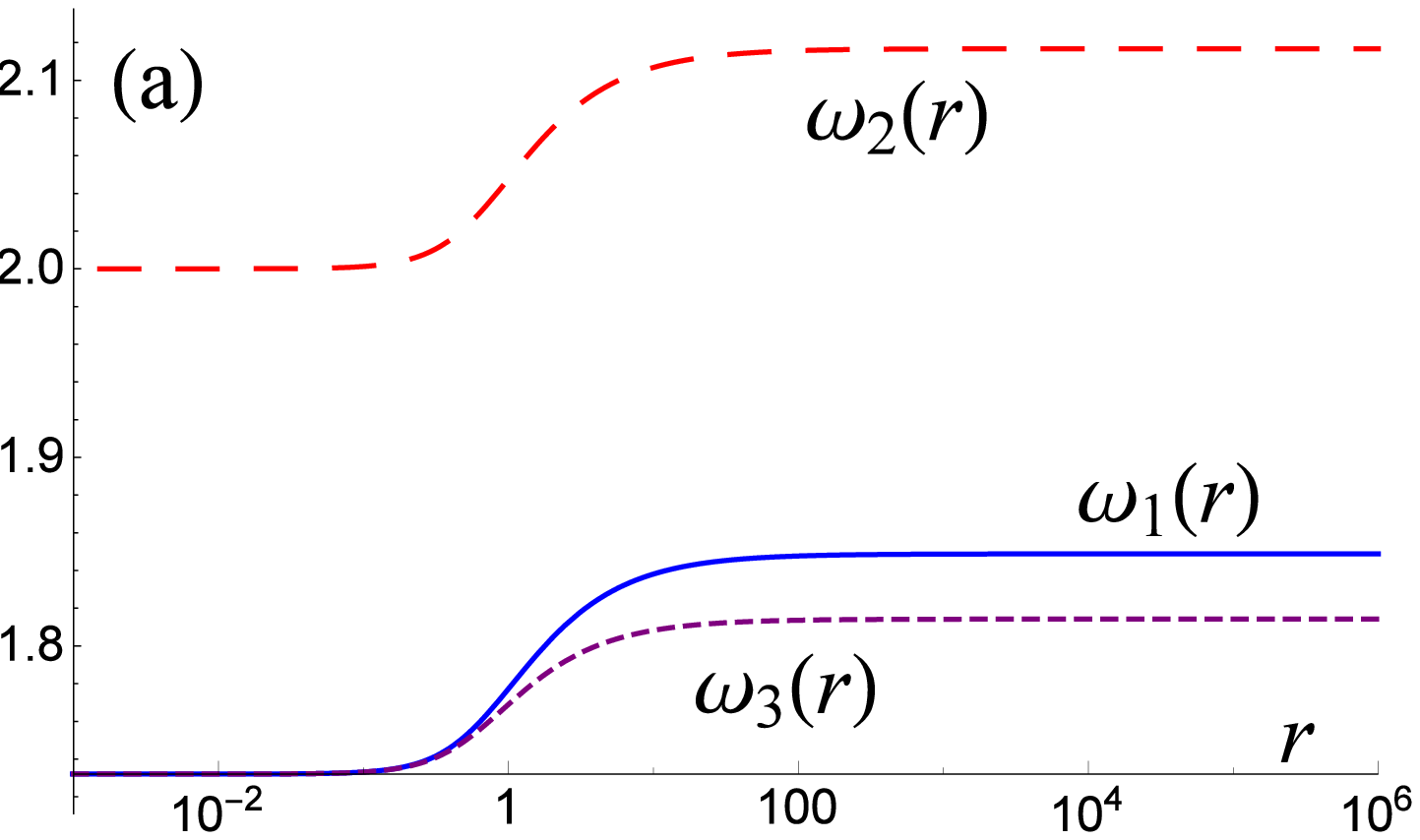}
\includegraphics[width=7.5cm]{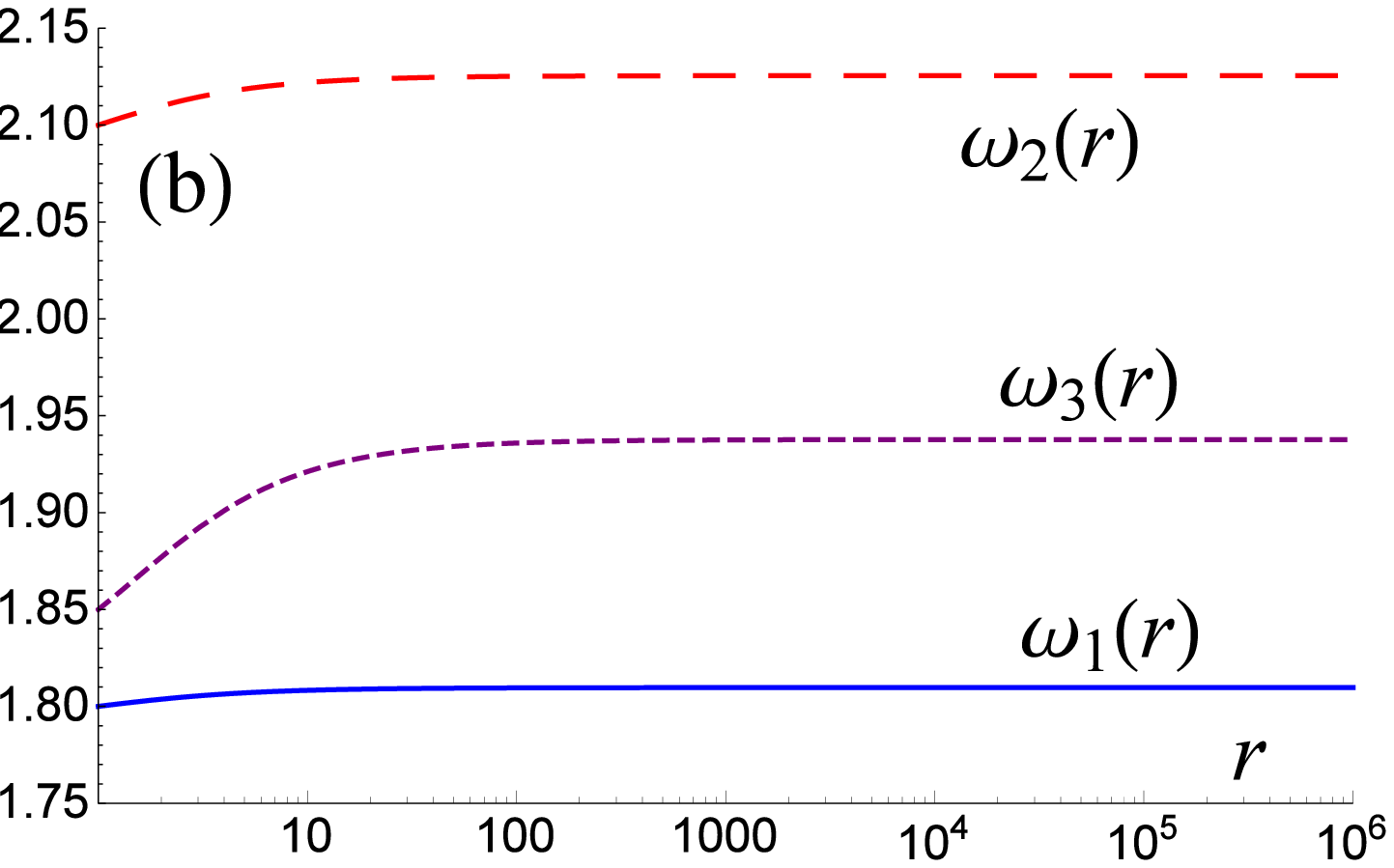}
\caption{Example nodeless solutions for ${\mathfrak {su}}(4)$ EYM with $\Lambda = -10$.  In each case we plot the gauge field functions $\omega _{1}(r)$, $\omega _{2}(r)$ and $\omega _{3}(r)$ (the typical behaviour of the metric functions can be found in the example solutions plotted in Ref.~\onlinecite{Baxter2}).
In (a) we show a soliton solution, and in (b) a black hole solution with $r_{h}=1$.}
\label{fig:2}
\end{figure}

In Ref.~\onlinecite{Baxter3} the existence of genuinely ${\mathfrak {su}}(N)$ solutions of the static field equations (\ref{eq:Ee}, \ref{eq:YMe}) in a neighbourhood of the above embedded ${\mathfrak {su}}(2)$ solutions was proven for all $N$, for sufficiently large $\left| \Lambda \right| $.
In this article we focus on those ${\mathfrak {su}}(N)$ solutions which are close to embedded ${\mathfrak {su}}(2)$ solutions and for which all the gauge field
functions $\omega _{j}(r)$ have no zeros.
In Refs.~\onlinecite{Baxter2, Baxter1} we have presented various phase space plots for ${\mathfrak {su}}(3)$ and ${\mathfrak {su}}(4)$
which demonstrate numerically the existence of regions of these nodeless soliton and black hole solutions.
Here we simply plot, in figures \ref{fig:1} and \ref{fig:2}, examples of nodeless soliton and black hole solutions for ${\mathfrak {su}}(3)$ and
${\mathfrak {su}}(4)$ respectively, referring the reader to Ref.~\onlinecite{Baxter2} for further details of the phase space of solutions.

\subsection{Perturbation equations}
\label{sec:perturbation}

In this paper we are interested in linear, spherically symmetric, perturbations of the static equilibrium solutions discussed in section \ref{sec:static}.
Our particular interest is in time-periodic, bound state, perturbations which vanish at either the origin or event horizon (as applicable, for soliton and
black hole solutions respectively) and at infinity.

To derive the perturbation equations, we write our time-dependent field variables as
a sum of the static equilibrium quantities (denoted by a bar, e.g.~${\ov {\mu }}(r)$) plus small perturbations
(denoted by a $\delta $, e.g.~$\delta \mu (t,r)$) as follows:
\begin{eqnarray}
\mu (t,r) & = & {\ov {\mu }}(r) + \delta \mu (t,r),
\nonumber \\
S(t,r) & = & {\ov {S}}(r) + \delta S(t,r),
\nonumber \\
m(t,r) & = & {\ov {m}}(r) + \delta m(t,r),
\nonumber \\
\Delta (t,r) & = & {\ov {\Delta }}(r) +\delta \Delta (t,r),
\nonumber \\
\alpha _{j} (t,r)  & = & \delta \alpha _{j} (t,r),
\nonumber \\
\beta _{j}(t,r) & = & \delta \beta _{j}(t,r),
\nonumber \\
\omega _{j}(t,r) & = & {\ov {\omega }}_{j}(r) + \delta \omega _{j}(t,r),
\nonumber \\
\gamma _{j}(t,r) & = & \delta \gamma _{j}(t,r).
\label{eq:perturbations}
\end{eqnarray}
Recall from section \ref{sec:static} that the gauge field functions $\alpha _{j}$,
$\beta _{j}$ and $\gamma _{j}$ all vanish for static equilibrium solutions, but here we consider non-zero perturbations of these parts of the gauge potential.

The perturbation equations are found by substituting the field variables in the form (\ref{eq:perturbations}) into the components of the Einstein tensor (\ref{eq:EinsteinTensor}) and gauge field (\ref{eq:Fmunu}), and then working out the field equations (\ref{eq:fieldeqns}).  We work only to first order in the perturbations and simplify the resulting equations using the static equilibrium field equations (\ref{eq:Ee}, \ref{eq:YMe}).

First of all, the linearized Einstein perturbation equations become
\begin{subequations}
\begin{eqnarray}
\delta \mu ' & = & \frac {1}{r} \left[
-\delta \mu - 2{\ov {\Gamma }}\delta \mu - 2{\ov {\mu }} \delta \Gamma
- 2r^{2} \delta \Pi
\right] ,
\label{eq:deltamup}
\\
\delta {\dot {\mu }} & = & -\frac {2{\ov {\mu }}}{r} \delta H ,
\label{eq:deltamudot}
\\
\delta \Delta ' & = & \frac {2}{r} \delta \Gamma ,
\label{eq:deltadeltap}
\end{eqnarray}
\end{subequations}
where
\begin{eqnarray}
\delta \Gamma  & = &
2 \sum _{j=1}^{N-1} {\ov {\omega }}_{j}' \delta \omega _{j}',
\nonumber \\
\delta \Pi & = &
\frac {1}{r^{4}} \sum _{j=1}^{N} \left[
{\ov {\omega }}_{j}^{2} - {\ov {\omega }}_{j-1}^{2} - N - 1 +2j
\right]
\left[
{\ov {\omega }}_{j} \delta \omega _{j} - {\ov {\omega }}_{j-1} \delta \omega _{j-1}
\right],
\nonumber \\
\delta H & = & \sum _{j=1}^{N-1} 2{\ov {\omega }}_{j}'\delta {\dot {\omega }}_{j},
\label{eq:deltaPi}
\end{eqnarray}
and we remind the reader that we are using a dot to denote $\partial /\partial t$
and a prime to denote $\partial /\partial r$.

Given the form of the elements of the matrix $C$ (\ref{eq:Celements}), it is useful to consider the following combinations of the perturbations
$\delta \omega _{j}(t,r)$
and $\delta \gamma _{j}(t,r)$, for $j=1,\ldots N-1$:
\begin{eqnarray}
\delta \psi _{j} (t,r) & = & \delta \omega _{j}(t,r) + i{\ov {\omega }}_{j}(r)
\delta \gamma _{j}(t,r) ,
\nonumber \\
\delta \psi _{j}^{*} (t,r) & = & \delta \omega _{j}(t,r) - i{\ov {\omega }}_{j}(r)
\delta \gamma _{j}(t,r) ,
\end{eqnarray}
in terms of which the entries of the matrix $C$ are, to first order in the perturbations:
\begin{equation}
C_{j,j+1} = {\ov {\omega }}_{j}(r) + \delta \psi _{j}(t,r).
\end{equation}

In terms of $\delta \psi _{j}$, $\delta \psi _{j}^{*}$, the linearized
Yang-Mills perturbation equations are:
\begin{widetext}
\begin{subequations}
\begin{eqnarray}
0 & = &
{\ov {\mu }}\left[
\delta {\dot{\beta}}_{j}'-\delta \alpha_j''
+\left( \delta{\dot{\beta}}_{j}-\delta \alpha_j' \right)
\left( \frac{2}{r}-\frac{{\ov {S}}'}{{\ov {S}}}\right)
\right]
+
\frac{1}{2r^2}
\left[{\ov {\omega}}_{j}
\left( \delta{\dot {\psi }}_j-\delta {\dot {\psi}}_j^* \right)
\right. \nonumber \\ & & \left.
-{\ov {\omega }}_{j-1}
\left( \delta {\dot {\psi }}_{j-1}-\delta {\dot {\psi }}_{j-1}^* \right)
+ 2{\ov {\omega }}_j^{2}
\left( \delta\alpha_j-\delta\alpha_{j+1} \right)
-2{\ov {\omega }}_{j-1}^{2}
\left( \delta\alpha_{j-1}-\delta\alpha_j \right) \right] ,
\label{eq:tYMpert}
\\
0 & = &
\frac{1}{{\ov {\mu }} {\ov {S}}^2}
\left(\delta {\ddot {\beta }}_j -\delta {\dot {\alpha }}_j' \right)
+\frac{1}{2r^2}
\left[
{\ov {\omega }}_j \left( \delta\psi_j' -\delta\psi_j^{*'} \right)
-{\ov {\omega}}_{j-1} \left( \delta\psi_{j-1}'-\delta\psi_{j-1}^{*'} \right)
-{\ov {\omega }}_j' \left(\delta\psi_j-\delta\psi_j^* \right)
\right. \nonumber \\ & & \left.
+{\ov {\omega }}_{j-1}' \left( \delta\psi_{j-1}-\delta\psi_{j-1}^* \right)
+2{\ov {\omega }}_{j-1}^{2} \left( \delta\beta_j-\delta\beta_{j-1} \right)
-2{\ov {\omega }}_j^{2}\left(\delta\beta_{j+1}-\delta\beta_j \right)
\right],
\label{eq:rYMpert}
\\
0 & = &
-\frac{1}{2{\ov {\mu }} {\ov {S}}^2}
\left[ \delta {\ddot {\psi }}_{j}
+{\ov {\omega }}_j \left( \delta {\dot {\alpha }}_j
-\delta {\dot {\alpha}}_{j+1} \right) \right]
+\frac {1}{2}{\ov {\omega }}_{j}'' \delta \mu
+\frac{{\ov {\mu }}}{2} \left[
\delta\psi_j''
+{\ov {\omega }}_j \left( \delta\beta_j'-\delta\beta_{j+1}' \right)
\right. \nonumber \\ & & \left.
+2{\ov {\omega}}_j' \left( \delta\beta_j-\delta\beta_{j+1} \right) \right]
+ \frac{{\ov {\omega }}_{j}'}{2}\left[
\delta\mu'+{\ov {\mu }} \delta \left(\frac{S'}{S}\right)
+ \frac{{\ov {S}}'}{{\ov {S}}}\delta\mu \right]
+ \frac {{\ov {\mu }}'}{2} \left[
\delta\psi'_j+{\ov {\omega }}_j \left(
\delta\beta_j-\delta\beta_{j+1} \right) \right]
\nonumber \\ & &
+\frac{1}{2r^2} \left[
-{\ov {\omega }}_j^{2} \left(\delta\psi_j +\delta\psi_j^* \right)
+\frac{1}{2}{\ov {\omega }}_{j}{\ov {\omega }}_{j+1}
\left( \delta\psi_{j+1}+\delta\psi_{j+1}^* \right)
+\frac{1}{2}{\ov {\omega }}_{j} {\ov {\omega }}_{j-1}
\left( \delta\psi_{j-1}+\delta\psi_{j-1}^* \right)
\right. \nonumber \\ & &  \left.
+W_{j} \delta\psi_j
\right] ,
\label{eq:qrealYMpert}
\\
0 & = &
-\frac{1}{2{\ov {\mu }} {\ov {S}}^2}
\left[
-\delta {\ddot {\psi }}_j^* +{\ov {\omega }}_j
\left( \delta {\dot {\alpha }}_j-\delta {\dot {\alpha }}_{j+1} \right)
\right]
-\frac {1}{2}{\ov {\omega }}_{j}'' \delta \mu
+\frac{{\ov {\mu }}}{2} \left[
-\delta\psi_j^{*''}
+{\ov {\omega }}_j \left( \delta\beta_j'-\delta\beta_{j+1}' \right)
\right. \nonumber \\ & &  \left.
+2{\ov {\omega }}_j' \left( \delta\beta_j-\delta\beta_{j+1} \right) \right]
-\frac{{\ov {\omega }}_{j}'}{2} \left[
\delta\mu'+{\ov {\mu }} \delta \left(\frac{S'}{S}\right)
+\frac{{\ov {S}}'}{{\ov {S}}} \delta\mu
\right]
+ \frac {{\ov {\mu }}'}{2} \left[
-\delta\psi_j^{*'}+{\ov {\omega }}_j \left( \delta\beta_j-\delta\beta_{j+1} \right) \right]
 \nonumber \\ & &
+\frac{1}{2r^2}\left[
{\ov {\omega }}_j^{2} \left( \delta\psi_j+\delta\psi_j^* \right)
-\frac{1}{2}{\ov {\omega }}_{j} {\ov {\omega }}_{j+1}
\left( \delta\psi_{j+1}+\delta\psi_{j+1}^* \right)
-\frac{1}{2}{\ov {\omega }}_{j} {\ov {\omega }}_{j-1}
\left( \delta\psi_{j-1}+\delta\psi_{j-1}^* \right)
\right. \nonumber \\ & &  \left.
- W_{j} \delta\psi_j^*
\right] .
\label{eq:qimYMpert}
\end{eqnarray}
\end{subequations}
\end{widetext}
The equations (\ref{eq:tYMpert}--\ref{eq:rYMpert}) come from the $t$ and $r$ Yang-Mills equations, respectively, and there are $N$ of each of these equations, corresponding to $j=1,\ldots N$.
The equations (\ref{eq:qrealYMpert}--\ref{eq:qimYMpert}) come from taking the real and imaginary parts of the $\theta $ Yang-Mills equation (the $\phi $ Yang-Mills equation gives the same pair of equations), assuming that all perturbations are real, and there are $N-1$ of each of these equations, corresponding to $j=1,\ldots N-1$.

Our time-dependent, spherically symmetric gauge field ansatz (\ref{eq:gaugepot}) has a residual gauge degree of freedom.  For a diagonal matrix ${\mathfrak {g}}(t,r)$, consider the following gauge transformation:
\begin{eqnarray}
{\mathcal {A}} & \rightarrow & {\mathcal {A}}+{\mathfrak {g}}^{-1}
{\dot{{\mathfrak {g}}}} ,
\nonumber \\
{\mathcal {B}} & \rightarrow & {\mathcal {B}}
+ {\mathfrak {g}}^{-1} {\mathfrak {g}}' ,
\nonumber \\
C-C^H & \rightarrow & {\mathfrak {g}}^{-1}\left( C-C^H \right) {\mathfrak {g}} ,
\nonumber \\
C+C^H & \rightarrow &
{\mathfrak {g}}^{-1}\left( C+C^H \right) {\mathfrak {g}} ,
\label{eq:gaugetrans}
\end{eqnarray}
under which the gauge field transforms as
\begin{equation}
F_{\mu \nu } \rightarrow {\mathfrak {g}}^{-1} F_{\mu \nu } {\mathfrak {g}}.
\end{equation}
We choose the diagonal matrix ${\mathfrak {g}}$ so that
${\mathcal {A}}+{\mathfrak {g}}^{-1} {\dot{{\mathfrak {g}}}} =0 $, which enables us to set the perturbations $\delta \alpha _{j}(t,r) \equiv 0$ for all $j=1,\ldots N$.

With this choice of gauge, the perturbation equations
(\ref{eq:deltamup}--\ref{eq:deltadeltap}, \ref{eq:tYMpert}--\ref{eq:qimYMpert})
decouple into two sectors.
The first sector contains the Yang-Mills perturbations $\delta \beta  _{j}$,
$j=1,\ldots , N$ and $\delta \gamma _{j}$, $j=1,\ldots , N-1$ and does not contain any metric perturbations.
This sector is known as the {\em {sphaleronic}} sector \cite{Lavrelashvili1}.
This terminology arises from the fact that the ${\mathfrak {su}}(2)$ EYM solitons \cite{Bartnik1} and black holes \cite{Bizon1} in asymptotically flat space
possess instabilities in this sector \cite{Lavrelashvili1,Volkov2,sphalstab} analogous to the unstable mode of the Yang-Mills-Higgs sphaleron \cite{sphaleron}.
The second sector contains the perturbations of the metric functions $\delta \mu $ and $\delta \Delta $ and the Yang-Mills perturbations $\delta \omega _{j}$,
$j=1,\ldots , N-1$.
This sector is known as the {\em {gravitational}} sector.
As the static equilibrium solutions are purely magnetic and spherically symmetric, they are invariant under a parity transformation.  As a result of this additional symmetry, the two decoupled sectors of perturbations transform in a particular way under a parity transformation: the perturbations in the sphaleronic sector have odd parity and change sign under a parity transformation; the perturbations in the gravitational sector have even parity and do not change under a parity transformation.

In the analysis of the sphaleronic and gravitational perturbation sectors in sections \ref{sec:sphaleronic} and \ref{sec:gravitational} respectively, we will change our independent radial variable to the usual `tortoise' co-ordinate $r_{*}$, defined by
\begin{equation}
\frac {dr_{*}}{dr} = \frac {1}{{\ov {\mu }}{\ov {S}}}.
\label{eq:tortoise}
\end{equation}
For perturbations of static soliton solutions, we choose the constant of integration
such that $r_{*}=0$ at the origin where $r=0$.  In this case $r_{*}$ has a maximum value, $r_{c}$, as $r\rightarrow \infty $.
For perturbations of static black hole solutions, we choose the constant of integration such that $r_{*} \rightarrow 0 $ as $r\rightarrow \infty $, and then $r_{*}\rightarrow -\infty $ as the event horizon is approached, $r\rightarrow r_{h}$.

\section{Sphaleronic sector perturbations}
\label{sec:sphaleronic}

The sphaleronic sector consists of the odd parity Yang-Mills perturbations
$\delta \beta _{j}$ ($j=1,\ldots , N$) and $\delta \gamma _{j}$ ($j=1,\ldots , N-1$).
In the gauge $\delta \alpha _{j} \equiv 0$, $j=1,\ldots , N$, the sphaleronic sector perturbation equations are (\ref{eq:tYMpert}, \ref{eq:rYMpert}) and a third perturbation equation which comes from adding equations (\ref{eq:qrealYMpert}) and
(\ref{eq:qimYMpert}). The equations are:
\begin{subequations}
\begin{eqnarray}
0 & = & {\ov {\mu }} \left[
\delta {\dot{\beta }}_j' + \delta {\dot{\beta }}_j
\left(
\frac{2}{r}-\frac{{\ov {S}}'}{{\ov {S}}}
\right) \right]
+\frac{1}{r^2} \left[
{\ov {\omega }}_j^{2} \delta {\dot{\gamma }}_{j}
-{\ov {\omega }}_{j-1}^{2} \delta {\dot {\gamma}}_{j-1}
\right] ,
\label{eq:sphalsector1}
\\
0 & = &
-\frac{1}{{\ov {\mu}} {\ov {S}}^2}
\delta {\ddot {\beta }}_j
+\frac{1}{r^2} \left[
-{\ov {\omega }}_j^{2} \delta\gamma_j'+{\ov {\omega }}_{j-1}^{2}\delta\gamma_{j-1}'
-{\ov {\omega }}_{j-1}^{2}
\left( \delta\beta_j-\delta\beta_{j-1} \right)
+{\ov {\omega }}_j^{2}\left( \delta\beta_{j+1}-\delta\beta_j \right) \right] ,
\nonumber \\ & &
\label{eq:sphalsector2}
\\
0 & = & -\frac{1}{{\ov {\mu }} {\ov {S}}^2}{\ov {\omega }}_j
\delta {\ddot {\gamma }}_j +{\ov {\mu }}{\ov {\omega }}_j\delta\gamma_j''
+{\ov {\mu}}{\ov {\omega }}_{j} \left( \delta\beta_j'-\delta\beta_{j+1}' \right)
\nonumber \\ & &
+\left[
2{\ov {\mu }}{\ov {\omega }}_j' +{\ov {\mu }}'{\ov {\omega }}_j
+{\ov {\mu }}{\ov {\omega }}_j \frac{{\ov {S}}'}{{\ov {S}}}\right]
\left[
\delta\gamma_j'+\delta\beta_j-\delta\beta_{j+1}\right] .
\label{eq:sphalsector3}
\end{eqnarray}
\end{subequations}

\subsection{Sphaleronic sector perturbation equations in matrix form}
\label{sec:smatrix}

We first simplify these equations by changing our radial co-ordinate from $r$ to the tortoise co-ordinate $r_{*}$  (\ref{eq:tortoise}) and by introducing new dependent
variables $\delta \epsilon _{j}$ ($j=1,\ldots , N$) and $\delta \Phi _{j}$ ($j=1,\ldots , N-1$) by:
\begin{equation}
\delta \epsilon _{j} = r{\sqrt {\ov {\mu }}} \, \delta \beta _{j},
\qquad
\delta \Phi _{j} = {\ov {\omega }}_{j} \delta \gamma _{j}.
\label{eq:newsvars}
\end{equation}
The perturbation equations (\ref{eq:sphalsector1}--\ref{eq:sphalsector3}) then take the form
\begin{subequations}
\begin{eqnarray}
0 & = &
\drs \delta {\dot {\epsilon }}_j
+
\left(
\frac{2{\ov {\mu }} {\ov {S}}}{r}
-\frac{\drs {\ov {S}}}{{\ov {S}}}
\right)
\delta {\dot {\epsilon }}_j
+\frac {{\ov {S}}{\sqrt {\ov {\mu }}}}{r}
\left(
{\ov {\omega }}_j \delta {\dot {\Phi }}_j
-{\ov {\omega }}_{j-1} \delta {\dot {\Phi}}_{j-1}
\right) ,
\label{eq:gaussc1}
\\
\delta {\ddot {\epsilon }}_j & = &
h
\left[
-{\ov {\omega }}_j \drs \delta \Phi_j
+ {\ov {\omega }}_{j-1} \drs \delta \Phi_{j-1}
+  \left( \drs {\ov {\omega }}_j \right) \delta \Phi_j
-\left( \drs {\ov {\omega }}_{j-1} \right)
\delta\Phi_{j-1}
\right]
\nonumber \\ & &
+h^{2}
\left[
{\ov {\omega }}_j^{2} \left( \delta \epsilon _{j+1}- \delta \epsilon _j \right)
-{\ov {\omega }}_{j-1}^{2} \left( \delta \epsilon _j - \delta \epsilon _{j-1} \right)
\right],
\\
\delta {\ddot {\Phi }}_j & = &
\drs ^{2}\delta \Phi_j
-\frac{\drs ^{2}{\ov {\omega }}_j}{{\ov {\omega }}_{j}} \delta \Phi _{j}
+ h{\ov {\omega }}_{j}
\drs \left( \delta \epsilon _j-\delta \epsilon _ {j+1} \right)
+
\left[ \drs \left( h {\ov {\omega }}_{j} \right) + h\drs {\ov {\omega }}_{j} \right]
\left( \delta \epsilon _j - \delta \epsilon _{j+1} \right) ,
\nonumber \\ &  &
\label{eq:sphsectorint3}
\end{eqnarray}
\end{subequations}
where we have introduced the quantity
\begin{equation}
h = \frac {{\ov {S}}{\sqrt {\ov {\mu }}}}{r}.
\end{equation}
We now express the perturbation equations in matrix form by defining
\begin{eqnarray}
{\bmath {\delta \epsilon }} & = & \left( \delta \epsilon _{1},\ldots ,
\delta \epsilon _{N}
\right) ^{T} ,
\nonumber \\
{\bmath {\delta \Phi }} & = & \left( \delta \Phi _{1},\ldots , \delta \Phi _{N-1}
\right) ^{T},
\end{eqnarray}
in terms of which the perturbation equations (\ref{eq:gaussc1}--\ref{eq:sphsectorint3}) take the form
\begin{subequations}
\begin{eqnarray}
0 & = &
\drs \left( h^{-1} {\bmath {\delta }}{\dot {\bmath {\epsilon }}} \right)
+ {\mathcal {F}} {\bmath {\delta }}{\dot {\bmath {\Phi }}} ,
\label{eq:gaussc2}
\\
{\bmath {\delta }} {\ddot {{\bmath {\epsilon }}}} & = &
h^{2} {\mathcal {K}} {\bmath {\delta \epsilon }}
- h \left[
{\mathcal {F}} \drs {\bmath {\delta \Phi }} - \left( \drs {\mathcal {F}} \right)
{\bmath {\delta \Phi }} \right] ,
\label{eq:sphsectorvector1}
\\
{\bmath {\delta }}{\ddot {{\bmath {\Phi }}}}
& = &
\drs ^{2} {\bmath {\delta \Phi }}
+ h {\mathcal {F}}^{T} \drs {\bmath {\delta \epsilon }}
+ {\mathcal {X}} {\bmath {\delta \epsilon }}
+ {\mathcal {W}} {\bmath {\delta \Phi }} ,
\label{eq:sphsectorvector2}
\end{eqnarray}
\end{subequations}
where we have defined an $N\times \left( N -1 \right) $ matrix ${\mathcal {F}}$,
an $N \times N$ matrix ${\mathcal {K}}$, an $\left( N -1 \right) \times \left(
N-1 \right)$ matrix ${\mathcal {W}}$ and an $\left( N-1 \right) \times N$ matrix
${\mathcal {X}}$ as follows:
\begin{subequations}
\begin{eqnarray}
{\mathcal {F}} & = &
\left(
\begin{array}{ccccc}
{\ov {\omega }}_1&0&0&\cdots&0\\
-{\ov {\omega}}_1&{\ov {\omega }}_2&0&\cdots&0\\
0&-{\ov {\omega }}_2&{\ov {\omega }}_3&\cdots&0\\
\vdots&\vdots&\vdots&\ddots&{\ov {\omega }}_{N-1}\\
0&0&0&0&-{\ov {\omega }}_{N-1}\\
\end{array}
\right) ,
\\
{\mathcal {K}} & = &
\left(
\begin{array}{ccccc}
-{\ov {\omega }}_1^{2}& {\ov {\omega }}_1^{2}&0&\cdots&0\\
{\ov {\omega }}_1^{2}&-{\ov {\omega }}_1^{2}-{\ov {\omega }}_2^{2}
& {\ov {\omega }}_2^{2}&\cdots&0\\
0&{\ov {\omega  }}_2^{2}&-{\ov {\omega }}_2^{2}
-{\ov {\omega }}_3^{2}&\cdots&0\\
\vdots&\vdots&\vdots&\ddots&{\ov {\omega }}_{N-1}^{2}\\
0&0&0&{\ov {\omega }}_{N-1}^{2}&- {\ov {\omega }}_{N-1}^{2}\\
\end{array}
\right) ,
\nonumber \\ & &
\label{eq:Kdef}
\\
{\mathcal {W}} & = &
h^{2}
\,
\Diag \left(
W_{1}, \ldots , W_{N-1} \right) ,
\label{eq:Wmatrixdef}
\\
{\mathcal {X}} & = &
2h\drs {\mathcal {F}}^{T}
+ \left( \drs  h \right) {\mathcal {F}}^{T},
\end{eqnarray}
\end{subequations}
where the quantities $W_{j}$, $j=1,\ldots , N-1$ are given by (\ref{eq:Wdef}).
Finally, we introduce a vector ${\bmath {\Psi }}$ of dimension $2N-1$ by
\begin{equation}
{\bmath {\Psi }} = \left(
\begin{array}{c}
{\bmath {\delta \epsilon }} \\
{\bmath {\delta \Phi }}
\end{array}
\right) ,
\end{equation}
in terms of which the first perturbation equation (\ref{eq:gaussc2}) takes the form
\begin{equation}
{\mathcal {G}} {\dot {{\bmath {\Psi }}}} \equiv
\drs
\left[ h^{-1}
\left(
\begin{array}{cc}
{\mathcal {I}}_{N} & 0 \\
0 & 0
\end{array}
\right)
{\dot {\bmath {\Psi }}}
\right]
+ \left(
\begin{array}{cc}
0 & {\mathcal {F}} \\
0 & 0
\end{array}
\right)
{\dot {{\bmath {\Psi }}}}
= 0,
\label{eq:gausscfinal}
\end{equation}
and the remaining equations (\ref{eq:sphsectorvector1}--\ref{eq:sphsectorvector2}) can be compactly written as
\begin{equation}
-{\ddot {{\bmath {\Psi }}}} =
{\mathcal {U}} {\bmath {\Psi }},
\label{eq:sphalpertfinal}
\end{equation}
where we have defined the operator
\begin{eqnarray}
{\mathcal {U}} {\bmath {\Psi }} & \equiv &
-\left(
\begin{array}{cc}
0 & 0 \\
0 & {\mathcal {I}}_{N-1}
\end{array}
\right)
\drs ^{2} {\bmath {\Psi }}
- h \left(
\begin{array}{cc}
0 & -{\mathcal {F}} \\
{\mathcal {F}}^{T} & 0
\end{array}
\right)
\drs {\bmath {\Psi }}
- \left(
\begin{array}{cc}
h^{2}{\mathcal {K}} &
h \drs {\mathcal {F}} \\
{\mathcal {X}} & {\mathcal {W}}
\end{array}
\right)
{\bmath {\Psi }}
\label{eq:Udef}
\end{eqnarray}
and ${\mathcal {I}}_{n}$ denotes the $n\times n$ identity matrix.

It is straightforward to show that the operator ${\mathcal {U}}$ (\ref{eq:Udef}) is
real and symmetric when acting on perturbations which vanish at either the origin or event horizon (as applicable) and at infinity.
However, as noted in Ref.~\onlinecite{Brodbeck}, the operator ${\mathcal {U}}$ is not elliptic and so the perturbation equation (\ref{eq:sphalpertfinal}) is not currently in hyperbolic form.

For time-periodic perturbations for which
${\bmath {\Psi }}(t,r)= e^{i\sigma t}{\bmath {\Psi }}(r)$, the perturbation equations (\ref{eq:sphalpertfinal}) take the form
\begin{equation}
\sigma ^{2}{\bmath {\Psi }} = {\mathcal {U}} {\bmath {\Psi  }}.
\label{eq:sphalperiodic}
\end{equation}
If we can show that the operator ${\mathcal {U}}$ is a positive operator, then the eigenvalues $\sigma ^{2}$ must also be positive.
This means that $\sigma $ is real and the perturbations are periodic in time.  In this case small perturbations remain small and there are
no unstable modes in the sphaleronic sector.
Our aim for the remainder of this section will be to show that there are at least some equilibrium ${\mathfrak {su}}(N)$ soliton and black
hole solutions for which ${\mathcal {U}}$ is a positive operator.

\subsection{The Gauss constraint}
\label{sec:Gaussc}

The first of the linearized Yang-Mills perturbation equations (\ref{eq:gausscfinal}) is known as the {\em {Gauss constraint}}.
A lengthy calculation reveals that the Gauss constraint propagates, in other words the perturbation equations (\ref{eq:sphsectorvector1}--\ref{eq:sphsectorvector2})
imply that
\begin{equation}
{\mathcal {G}} {\ddot {\bmath {\Psi }}} =0
\end{equation}
independently of the Gauss constraint.
Equivalently, we may write \cite{Brodbeck}
\begin{equation}
{\mathcal {G}}{\mathcal {U}} =0.
\label{eq:gaussprop}
\end{equation}
Following Ref.~\onlinecite{Brodbeck}, we integrate (\ref{eq:gausscfinal}) with respect to time, and choose the constant of integration (in this case an arbitrary function of $r$) so that:
\begin{equation}
 {\mathcal {G}} {\bmath {\Psi }} =0 ,
\label{eq:gausscstrong}
\end{equation}
which we will call the {\em {strong Gauss constraint}} \cite{Brodbeck}.
Suppose we have a vector of perturbations ${\bmath {\Psi }}$ which satisfy the strong Gauss constraint at initial time $t=0$, and which initially satisfy the Gauss constraint (\ref{eq:gausscfinal}).
By virtue of (\ref{eq:gaussprop}), this vector of perturbations will satisfy the strong Gauss constraint at all subsequent times.

Now consider any vector of perturbations ${\bmath {\Psi }}$ (satisfying the perturbation equations) and write it as the sum of two parts: the first, ${\bmath {\Psi }}_{1}$, satisfying the strong Gauss constraint and the second, ${\bmath {\Psi }}_{2}$, failing to satisfy the strong Gauss constraint.
It is shown in Ref.~\onlinecite{Brodbeck} that the second vector of perturbations, ${\bmath {\Psi }}_{2}$,
is  pure gauge, having the form
\begin{equation}
{\bmath {\Psi }}_{2} = {\mathcal {G}}^{\dagger } {\bmath {\Upsilon}}
\end{equation}
where
\begin{equation}
{\mathcal {G}}^{\dagger }= -h^{-1}\left(
\begin{array}{cc}
{\mathcal {I}}_{N} & 0 \\
0 & 0
\end{array}
\right)
\drs +
\left(
\begin{array}{cc}
0 & 0 \\
{\mathcal {F}}^{T} & 0
\end{array}
\right)
\end{equation}
is the adjoint of the operator ${\mathcal {G}}$ (\ref{eq:gausscfinal}).
Such perturbations correspond to infinitesimal gauge transformations of the form (\ref{eq:gaugetrans}) with
(for small $\varepsilon $)
\begin{equation}
{\mathfrak {g}}=\exp \left( - \varepsilon {\tilde {{\bmath {\Upsilon}}}} \right)
\label{eq:gSGC}
\end{equation}
where ${\tilde {{\bmath {\Upsilon }}}}$ is an $N\times N$ matrix of the form
\begin{equation}
{\tilde {\bmath {\Upsilon}}} = \Diag
\left(
\Upsilon _{1}, \ldots , \Upsilon _{N}
\right)
\end{equation}
with $\Upsilon _{j}$, $j=1,\ldots, N$ the first $N$ elements in
${\bmath {\Upsilon }}$.
Therefore a vector of perturbations ${\bmath {\Psi }}$ which satisfy the Gauss constraint but not the strong Gauss constraint can be gauge-transformed
to a vector of perturbations satisfying the strong Gauss constraint.
Without loss of generality, we may therefore restrict attention to physical perturbations satisfying the strong Gauss constraint, which is essentially an initial condition.
Since this is a gauge transformation of initial data only, the matrix ${\mathfrak {g}}$ (\ref{eq:gSGC}) is time-independent, and so, by (\ref{eq:gaugetrans}),
this transformation preserves the gauge condition $\delta \alpha _{j}\equiv 0$.

\subsection{An alternative form of the operator ${\mathcal {U}}$}
\label{sec:altU}

In order to prove the existence of static solutions which have no unstable modes in the sphaleronic sector governed by the equations (\ref{eq:sphalpertfinal}), in the next subsection we will want to show that the symmetric operator ${\mathcal {U}}$ (\ref{eq:Udef}) is positive.
In this subsection we will write the operator in an alternative form which will enable us to find static equilibrium solutions for which
${\mathcal {U}}$ is a positive operator.

In particular, following Ref.~\onlinecite{Brodbeck}, we seek operators $\chi $ and ${\mathcal {V}}$ such that we may write
\begin{equation}
{\mathcal {U}} =  \chi ^{\dagger } \chi + {\mathcal {V}}
- {\mathcal {G}}^{\dagger } h^{2} {\mathcal {G}},
\label{eq:Vdef}
\end{equation}
where the operator ${\mathcal {G}}$ is given by (\ref{eq:gausscfinal}).
We then find
\begin{eqnarray}
{\mathcal {G}}^{\dagger }h^{2} {\mathcal {G}}
 & = &
-\left(
\begin{array}{cc}
{\mathcal {I}}_{N} & 0 \\
0 & 0
\end{array}
\right)
\drs ^{2}
+ h\left(
\begin{array}{cc}
0 & -{\mathcal {F}} \\
{\mathcal {F}}^{T} & 0
\end{array}
\right)
+
\left(
\begin{array}{cc}
h^{-1} \drs ^{2} h &
-\drs \left( h{\mathcal {F}} \right) - \left( \drs h \right) {\mathcal {F}}
\\
-\left( \drs h \right) {\mathcal {F}}^{T} &
h^{2} {\mathcal {F}}^{T} {\mathcal {F}}
\end{array}
\right) .
\nonumber \\ & &
\label{eq:GdaggerG}
\end{eqnarray}
Next define the operator $\chi $ as
\begin{equation}
\chi = \drs + h{\mathcal {Z}} ,
\label{eq:chidef}
\end{equation}
where ${\mathcal {Z}}$ is some $\left( 2N-1 \right) \times \left( 2N-1 \right) $ matrix which does not contain any derivative operators and which is to be determined.
Then
\begin{equation}
\chi ^{\dagger } \chi
= -\drs ^{2} + h\left( {\mathcal {Z}}^{T} - {\mathcal {Z}} \right) \drs
+\left[ h^{2} {\mathcal {Z}}^{T} {\mathcal {Z}} - \drs
\left( h{\mathcal {Z}} \right) \right] .
\label{eq:chidaggerchi}
\end{equation}
From now on we assume that ${\mathcal {Z}}$ is symmetric, so that there is no
first order derivative operator in $\chi ^{\dagger }\chi $.
Writing the matrix ${\mathcal {Z}}$ in the form
\begin{equation}
{\mathcal {Z}} = \left(
\begin{array}{cc}
{\mathcal {Z}}_{11} & {\mathcal {Z}}_{12} \\
{\mathcal {Z}}_{12}^{T} & {\mathcal {Z}}_{22}
\end{array}
\right) ,
\end{equation}
where ${\mathcal {Z}}_{11}$ is a symmetric $N \times N $ matrix,
${\mathcal {Z}}_{12}$ is an $N\times \left( N -1 \right)$ matrix and
${\mathcal {Z}}_{22}$ is a symmetric $\left( N-1 \right) \times \left( N-1 \right) $
matrix, using (\ref{eq:GdaggerG}, \ref{eq:chidaggerchi}) we find that the matrix
${\mathcal {V}}$ defined in (\ref{eq:Vdef}) has the form
\begin{equation}
{\mathcal {V}} = \left(
\begin{array}{cc}
{\mathcal {V}}_{11} & {\mathcal {V}}_{12} \\
{\mathcal {V}}_{21} & {\mathcal {V}}_{22}
\end{array}
\right) ,
\end{equation}
where
\begin{eqnarray}
{\mathcal {V}}_{11} & = &
-h^{2} {\mathcal {K}} + h^{2} \left( {\mathcal {Z}}_{11}^{T} {\mathcal {Z}}_{11} + {\mathcal {Z}}_{12} {\mathcal {Z}}_{12}^{T}  \right)
-\drs \left( h{\mathcal {Z}}_{11} \right)
+ h^{-1} \left(
\drs ^{2} h \right) {\mathcal {I}}_{N} ,
\nonumber \\
{\mathcal {V}}_{12} & = &
-2h\drs {\mathcal {F}} + h^{2} \left( {\mathcal {Z}}_{11}^{T} {\mathcal {Z}}_{12}
+{\mathcal {Z}}_{12} {\mathcal {Z}}_{22}^{T} \right)
-\drs \left( h {\mathcal {Z}}_{12} \right)
-2\left( \drs h \right) {\mathcal {F}},
\nonumber \\
{\mathcal {V}}_{21} & = &
-2\left( \drs  h \right) {\mathcal {F}}^{T}  - 2h \drs {\mathcal {F}}^{T}
+h^{2}\left( {\mathcal {Z}}_{12}^{T} {\mathcal {Z}}_{11}
+ {\mathcal {Z}}_{22}^{T} {\mathcal {Z}}_{12}^{T} \right)
- \drs \left( h{\mathcal {Z}}_{12}^{T} \right) ,
\nonumber \\
{\mathcal {V}}_{22} & = &
-{\mathcal {W}} + h^{2} \left( {\mathcal {Z}}_{12}^{T} {\mathcal {Z}}_{12}
+ {\mathcal {Z}}_{22}^{T} {\mathcal {Z}}_{22} \right)
-\drs \left( h{\mathcal {Z}}_{22} \right)
 + h^{2} {\mathcal {F}}^{T} {\mathcal {F}}.
\end{eqnarray}
We are free to choose the matrix ${\mathcal {Z}}$ so as to simplify the form of
${\mathcal {V}}$.
We first make the choices
\begin{equation}
{\mathcal {Z}}_{11}=h^{-2} \left( \drs h \right) {\mathcal {I}}_{N},
\qquad
{\mathcal {Z}}_{22}=0.
\end{equation}
In this case the form of ${\mathcal {V}}_{12}$ simplifies to
\begin{equation}
{\mathcal {V}}_{12} =
 -2\drs \left( h{\mathcal {F}} \right) - h\drs {\mathcal {Z}}_{12} ,
\end{equation}
which vanishes if we choose ${\mathcal {Z}}_{12}$ such that
\begin{equation}
{\mathcal {Z}}_{12} = -2 \int _{r_{*}=r_{*,{\mathrm {min}}}}^{r_{*}}
h^{-1} \drs \left( h{\mathcal {F}} \right) \, dr_{*} ,
\end{equation}
where $r_{*,{\mathrm {min}}}=0$ for equilibrium static soliton solutions and
$r_{*,{\mathrm {min}}}=-\infty $ for equilibrium static black hole solutions.
With this choice of ${\mathcal {Z}}_{12}$, it is straightforward to see that
${\mathcal {V}}_{21}$ also vanishes.

The matrix ${\mathcal {V}}$ is then block diagonal, with its diagonal entries being
\begin{eqnarray}
{\mathcal {V}}_{11} & = &
-h^{2} {\mathcal {K}} + h^{2} {\mathcal {Z}}_{12} {\mathcal {Z}}_{12}^{T}
 + \left( h^{-1} \drs h  \right) ^{2} {\mathcal {I}}_{N},
\nonumber \\
{\mathcal {V}}_{22} & = &
-{\mathcal {W}} + h^{2} {\mathcal {Z}}_{12}^{T} {\mathcal {Z}}_{12}
+ h^{2} {\mathcal {F}}^{T} {\mathcal {F}}.
\label{eq:Ventries}
\end{eqnarray}

\subsection{Conditions for no instabilities in the sphaleronic sector}
\label{sec:sphalstabconds}

For physical perturbations satisfying the strong Gauss constraint (\ref{eq:gausscstrong}), the
operator ${\mathcal {U}}$ (\ref{eq:Vdef}) appearing in the sphaleronic sector perturbation equations (\ref{eq:sphalperiodic}) reduces to
\begin{equation}
{\mathcal {U}} =  \chi ^{\dagger } \chi + {\mathcal {V}}.
\label{eq:Ureduced}
\end{equation}
Since the matrix ${\mathcal {V}}$ is symmetric, the operator ${\mathcal {U}}$ is symmetric and real.
From the form of the operator $\chi ^{\dagger} \chi $ (\ref{eq:chidaggerchi}), the operator (\ref{eq:Ureduced}) is elliptic.
Furthermore, since $\chi ^{\dagger }\chi $ is a positive operator, to show that ${\mathcal {U}}$ is a positive operator it suffices to show that ${\mathcal {V}}$ is a positive matrix.
The matrix ${\mathcal {V}}$ is block diagonal, and hence positive if its two non-zero diagonal blocks ${\mathcal {V}}_{11}$ and ${\mathcal {V}}_{22}$
(\ref{eq:Ventries}) are positive.

Let us begin with ${\mathcal {V}}_{11}$. The second and third terms in
${\mathcal {V}}_{11}$ are manifestly positive, so it remains to consider the term $-h^{2} {\mathcal {K}}$ where the matrix ${\mathcal {K}}$ is given by
(\ref{eq:Kdef}).
For an arbitrary vector ${\bmath {x}}=\left( x_{1},\ldots , x_{N} \right) ^{T}$, we have
\begin{eqnarray}
-{\bmath {x}}^{T} {\mathcal {K}} {\bmath {x}} & = &
{\ov {\omega }}_{1}^{2} x_{1}^{2}
+ \left( {\ov {\omega }}_{1}^{2} + {\ov {\omega }}_{2}^{2} \right) x_{2}^{2}
+ \ldots
+
\left( {\ov {\omega }}_{N-2}^{2} + {\ov {\omega }}_{N-1}^{2} \right) x_{N-1}^{2}
+ {\ov {\omega }}_{N-1}^{2} x_{N}^{2}
\nonumber \\ & &
- 2{\ov {\omega }}_{1}^{2}  x_{1}x_{2} - 2{\ov {\omega }}_{2}^{2} x_{2}x_{3} - \ldots
- 2{\ov {\omega }}_{N-1}^{2} x_{N-1}x_{N}
\nonumber \\ & = &
{\ov {\omega }}_{1}^{2} \left( x_{1}-x_{2} \right) ^{2}
+ {\ov {\omega }}_{2}^{2} \left( x_{2}-x_{3} \right) ^{2}
+ \ldots
 +
{\ov {\omega }}_{N-1}^{2} \left( x_{N-1}-x_{N} \right) ^{2}
\nonumber \\
& \ge & 0.
\end{eqnarray}
Therefore ${\mathcal {V}}_{11}$ is positive.

For ${\mathcal {V}}_{22}$, again the second and third terms are manifestly positive.
The first term, $-{\mathcal {W}}$ (\ref{eq:Wmatrixdef}) is a diagonal matrix, which will be positive if and only if its entries are positive.
For this to be the case, we require $W_{j}\le 0$ for $j=1,\ldots, N-1$, where
the quantities $W_{j}$ are defined in (\ref{eq:Wdef}).
This gives the following set of inequalities to be satisfied by the static equilibrium solutions for all $r$:
\begin{eqnarray}
{\ov {\omega }}_{1}^{2} & \ge &
1 + \frac {1}{2} {\ov {\omega }}_{2}^{2} ,
\nonumber \\
{\ov {\omega }}_{2}^{2} & \ge &
1 + \frac {1}{2} \left( {\ov {\omega }}_{1}^{2} + {\ov {\omega }}_{3}^{2} \right) ,
\nonumber \\
\vdots & &
\nonumber \\
{\ov {\omega }}_{j}^{2} & \ge &
1 + \frac {1}{2} \left( {\ov {\omega }}_{j-1}^{2} + {\ov {\omega }}_{j+1}^{2}
\right) ,
\nonumber \\
\vdots & &
\nonumber \\
{\ov {\omega }}_{N-1}^{2} & \ge &
1+\frac {1}{2} {\ov {\omega }}_{N-2}^{2}.
\label{eq:sstabineqs}
\end{eqnarray}
If the inequalities (\ref{eq:sstabineqs}) are satisfied for all $r$, then we can deduce that ${\mathcal {V}}_{22}$ is positive.
Therefore the matrix ${\mathcal {V}}$ is positive, and hence the operator ${\mathcal {U}}$ is a positive operator.
We deduce that physical solutions of the sphaleronic sector perturbation equations (\ref{eq:sphalperiodic}) must have $\sigma ^{2}$ positive,
so $\sigma $ is real and the perturbations are periodic in time.
Therefore small perturbations remain small and the equilibrium solutions have no instabilities in the sphaleronic sector.

We emphasize that the inequalities (\ref{eq:sstabineqs}) are sufficient for an equilibrium solution to have no unstable modes in the sphaleronic sector;
we have no expectation that these inequalities are necessary for stability.
Our interest in this paper is in proving the existence of stable soliton and black hole solutions of the ${\mathfrak {su}}(N)$ EYM equations.
Therefore we will have achieved this aim, at least for the sphaleronic sector, if we can find equilibrium solutions satisfying (\ref{eq:sstabineqs}) for all $r$.

\subsection{Special cases}
\label{sec:Sspecial}

Before proving the existence of non-trivial ${\mathfrak {su}}(N)$ equilibrium solutions which have no instabilities in the sphaleronic sector,
in this subsection we consider the embedded solutions discussed in section \ref{sec:embedded}.

\subsubsection{Schwarzschild-adS}
\label{sec:sSadS}

Setting ${\ov {\omega }}_{j} \equiv {\sqrt {j\left( N- j\right)}}$ and $m(r) \equiv M$,
we find that $W_{j}\equiv 0$ for all $j=1,\ldots , N-1$ and so the matrix
${\mathcal {W}}$ vanishes identically.
In this case the matrix ${\mathcal {V}}$ is manifestly positive and the operator
${\mathcal {U}}$ is positive when acting on physical perturbations.
Therefore the Schwarzschild-adS solution, as expected, has no instabilities in the sphaleronic sector.

\subsubsection{Reissner-Nordstr\"om-adS}
\label{sec:sRNadS}

If ${\ov {\omega }}_{j} \equiv 0$ for all $j=1,\ldots ,N-1$, then the sphaleronic sector perturbation equations reduce greatly. The perturbations $\delta \Phi _{j}$
(\ref{eq:newsvars}) vanish identically, leaving only the $\delta \epsilon _{j}$ perturbations. The only $\delta \epsilon _{j}$ perturbations which then satisfy the strong Gauss constraint have the form
\begin{equation}
\delta \epsilon _{j}= z_{j} h = z_{j} \frac {{\ov {S}}{\sqrt {\ov {\mu }}}}{r},
\end{equation}
where $z_{j}$ are arbitrary constants.
These perturbations do not vanish at the origin or the event horizon or infinity unless $z_{j}=0$ for all
$j=1,\ldots, N$.
For bound state perturbations which vanish at either the origin or the event horizon (as applicable) and at infinity, the only possibility is
$z_{j}=0$ and hence $\delta \epsilon _{j}\equiv 0$.
This means that there is no dynamics in the sphaleronic sector when the static equilibrium solution is embedded, magnetically-charged, Abelian
Reissner-Nordstr\"om-adS.  The only allowed perturbations of the gauge potential correspond to gauge transformations.

\subsubsection{Embedded ${\mathfrak {su}}(2)$ solutions}
\label{sec:ssu2}

Setting ${\ov {\omega }}_{j} \equiv {\ov {\omega }}(r){\sqrt {j\left( N-j\right) }}$,
the sphaleronic sector perturbation equations simplify considerably.
In particular, the matrix ${\mathcal {W}}$ (\ref{eq:Wmatrixdef}) reduces to
\begin{equation}
{\mathcal {W}} = h^{2} \left( 1- {\ov {\omega }}^{2} \right) {\mathcal {I}}_{N-1}.
\end{equation}
The inequalities (\ref{eq:sstabineqs}) are then all satisfied if
\begin{equation}
{\ov {\omega }}(r)^{2} \ge 1
\label{eq:su2sstabineqs}
\end{equation}
for all $r$, in which case the operator ${\mathcal {U}}$ is positive and there are no instabilities in the sphaleronic sector.

It remains to prove the existence of embedded ${\mathfrak {su}}(2)$ solutions satisfying (\ref{eq:su2sstabineqs}) for all $r$.
Embedded ${\mathfrak {su}}(2)$ black holes are parameterized by the radius of the event horizon $r_{h}$, the cosmological constant $\Lambda $ and
${\ov {\omega }}(r_{h})$.
Fix $r_{h}$ and choose ${\ov {\omega }}(r_{h})>1$.
From the Yang-Mills equation (\ref{eq:YMe}) in the ${\mathfrak {su}}(2)$ case, we have ${\ov {\omega }}'(r_{h})>0$, so that ${\ov {\omega }}$ is an increasing
function of $r$ close to the horizon.
Embedded ${\mathfrak {su}}(2)$ solitons are described by a single parameter $b$, such that, near the origin,
\begin{equation}
{\ov {\omega }}(r) = 1 + br^{2} + O(r^{3}),
\end{equation}
where, without loss of generality, we are assuming that ${\ov {\omega }}(0)>0$.
Choose $b>0$, so that ${\ov {\omega }}(r)>1$ in a neighbourhood of the origin and ${\ov {\omega }}'(r)>0$ for $r$ sufficiently small.
Also from (\ref{eq:YMe}), we see that the gauge function ${\ov {\omega }}$ cannot have a maximum if ${\ov {\omega }}>1$.
Therefore ${\ov {\omega }}$ will be an increasing function of $r$ for all $r\ge r_{h}$ for our black hole solution and all $r>0$ for the soliton solution.
Therefore (\ref{eq:su2sstabineqs}) will be satisfied and these solutions will have no instabilities in the sphaleronic sector.
The same argument applies if ${\ov {\omega }}<-1$ either at the horizon or near the origin.

In Ref.~\onlinecite{Winstanley1} it is proven that ${\mathfrak {su}}(2)$ black holes have no instabilities in the sphaleronic sector of ${\mathfrak {su}}(2)$ EYM
perturbations as long as the gauge function $\omega (r)$ has no zeros, with no further conditions on $\omega (r)$.
While the proof in Ref.~\onlinecite{Winstanley1} is for black holes only, the argument carries over trivially to the soliton case.
We note that here, we have a stronger sufficient condition (\ref{eq:su2sstabineqs}) for ${\mathfrak {su}}(2)$ solutions embedded in ${\mathfrak {su}}(N)$
EYM to have no instabilities in the ${\mathfrak {su}}(N)$ sphaleronic sector.
This is to be expected since the ${\mathfrak {su}}(N)$ EYM sphaleronic sector has more degrees of freedom ($2N-2$, comprising
$N$ functions $\beta _{j}$ whose sum must vanish and $N-1$ functions $\gamma _{j}$) than the ${\mathfrak {su}}(2)$
EYM sphaleronic sector (which has just two).

\subsection{Existence of static solutions with no sphaleronic sector instabilities}
\label{sec:sphstab}

We now turn to proving the existence of non-trivial ${\mathfrak {su}}(N)$ EYM solitons and black holes having no instabilities in the sphaleronic sector.
From the above analysis, all that is required is to show the existence of equilibrium solutions satisfying the inequalities (\ref{eq:sstabineqs}).
The argument for both soliton and black hole solutions is straightforward, based on results from Ref.~\onlinecite{Baxter3}.

Due to the symmetry of the field equations (\ref{eq:Ee}, \ref{eq:YMe}) under the transformation (\ref{eq:omegatransform}) it is sufficient to
consider gauge field functions such that $\omega _{j}>0$ near either the origin if we are considering a soliton solution or the event horizon if we are considering a black hole solution.
First we define the open region ${\mathcal {R}}$ which is the set of all positive values of the equilibrium
gauge field functions ${\ov {\omega }}_{j}>0$, $j=1,\ldots ,N-1$
such that $W_{j}<0$ (so that the inequalities (\ref{eq:sstabineqs}) are strictly satisfied for all points in ${\mathcal {R}}$).
From the argument in the previous subsection, there are embedded ${\mathfrak {su}}(2)$ soliton and black hole solutions such that the values of the gauge
field functions lie in ${\mathcal {R}}$ for all $r$.
From Proposition 9 of Ref.~\onlinecite{Baxter3}, there are genuinely (that is, non-embedded) ${\mathfrak {su}}(N)$ soliton and black hole solutions whose initial parameters (either near the origin or the event horizon, see section \ref{sec:boundary} for details)
lie in a neighbourhood of the initial parameters for the embedded ${\mathfrak {su}}(2)$ solitons and black holes.
Propositions 3 and 6 of Ref.~\onlinecite{Baxter3} tell us that the equilibrium gauge field functions ${\ov {\omega }}_{j}$ are analytic functions of the initial parameters and the radial co-ordinate.
Fix $r_{h}$ for the black hole solutions under consideration and set $r_{1}\gg \max \{ 1,r_{h} \}$ (with $r_{h}=0$ for soliton solutions).
Then, by analyticity, providing our ${\mathfrak {su}}(N)$ solutions have initial parameters sufficiently close to the initial parameters for the embedded
${\mathfrak {su}}(2)$ solutions, the gauge field functions ${\ov {\omega }}_{j}(r)$ for the ${\mathfrak {su}}(N)$ solutions will remain close to the embedded ${\mathfrak {su}}(2)$ solutions for all $r\le r_{1}$  and hence also within the
region ${\mathcal {R}}$ for all $r\le r_{1}$.
Providing we have chosen $r_{1}$ sufficiently large, for $r>r_{1}$ we are in the asymptotic large $r$ regime discussed in section 4.2 of Ref.~\onlinecite{Baxter3}.
The upshot of that analysis is that, by taking $r_{1}$ sufficiently large, the change in the gauge field functions as $r\rightarrow \infty $ from $r=r_{1}$
can be made arbitrarily small.
Therefore, since ${\mathcal {R}}$ is an open region, the gauge field functions ${\ov {\omega }}_{j}$ for our ${\mathfrak {su}}(N)$ solutions will remain
inside ${\mathcal {R}}$ for all $r>r_{1}$.

\begin{figure}
\includegraphics[width=7.5cm]{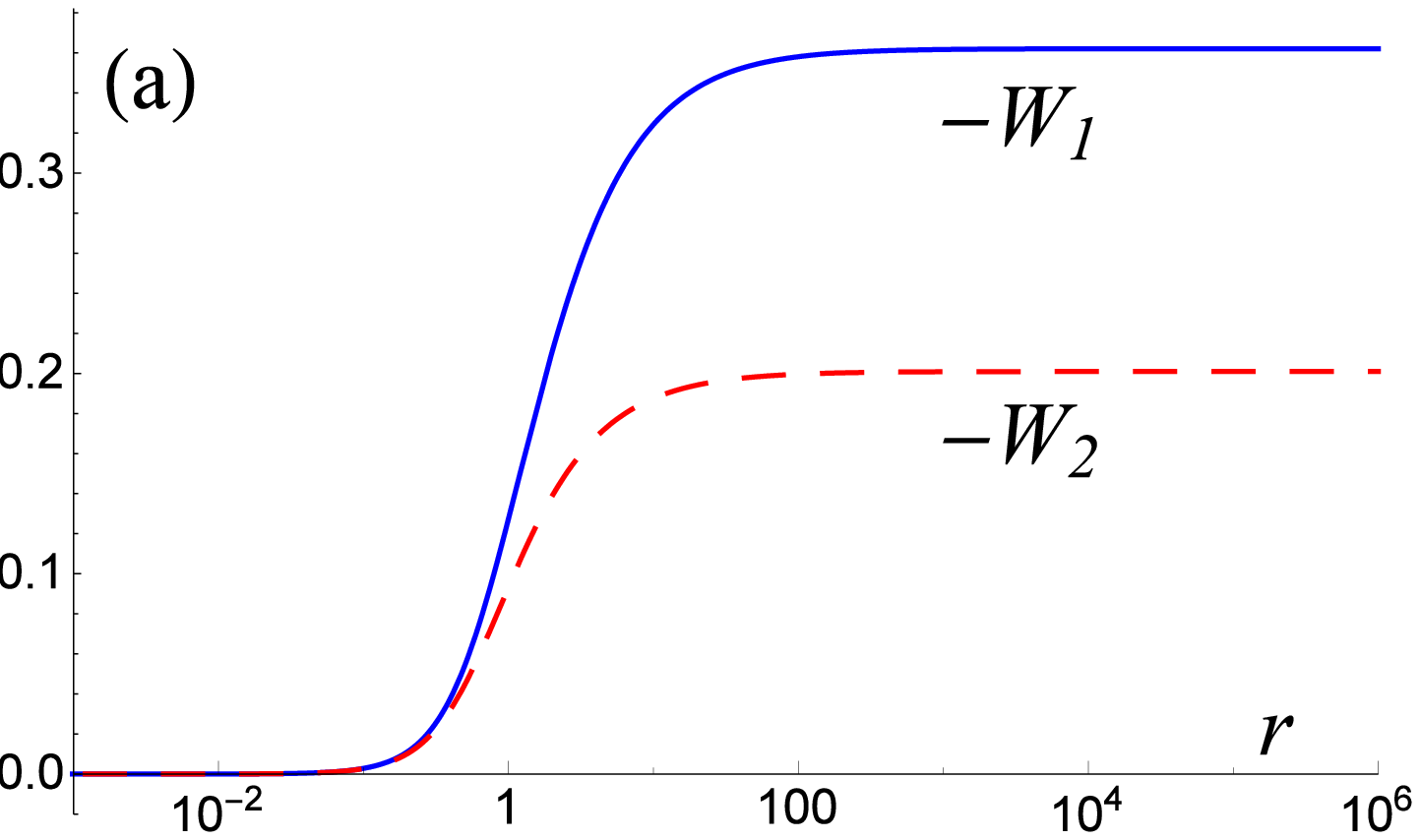}
\includegraphics[width=7.5cm]{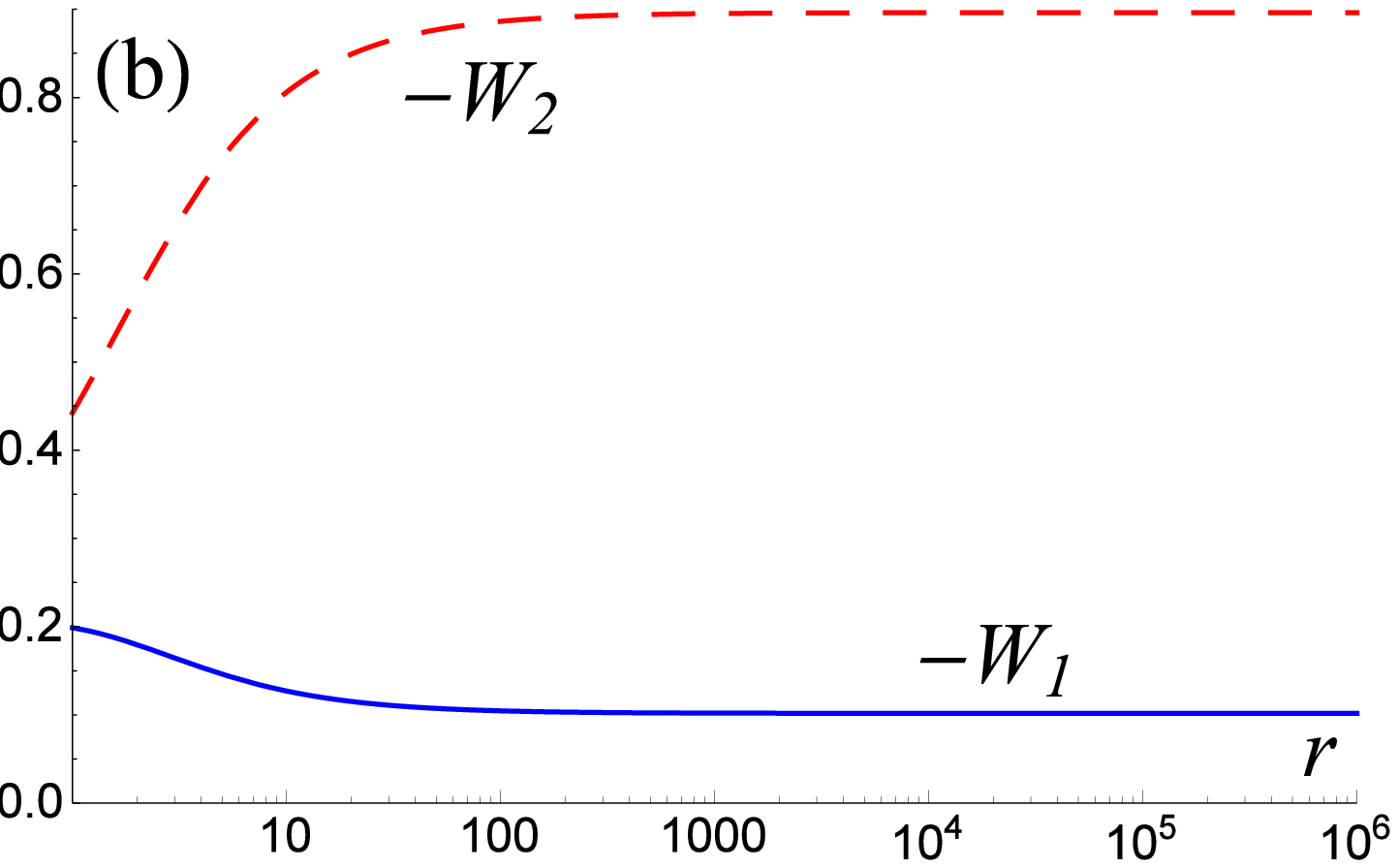}
\caption{$-W_{1}(r)$ and $-W_{2}(r)$ for the example nodeless  ${\mathfrak {su}}(3)$ solutions shown in figure \ref{fig:1}, (a) soliton and (b) black hole.
In both cases $-W_{j}\ge 0$ for all $r$, so that the inequalities (\ref{eq:sstabineqs}) are satisfied and the solutions have no instabilities in the
sphaleronic sector.}
\label{fig:3}
\end{figure}

\begin{figure}
\includegraphics[width=7.5cm]{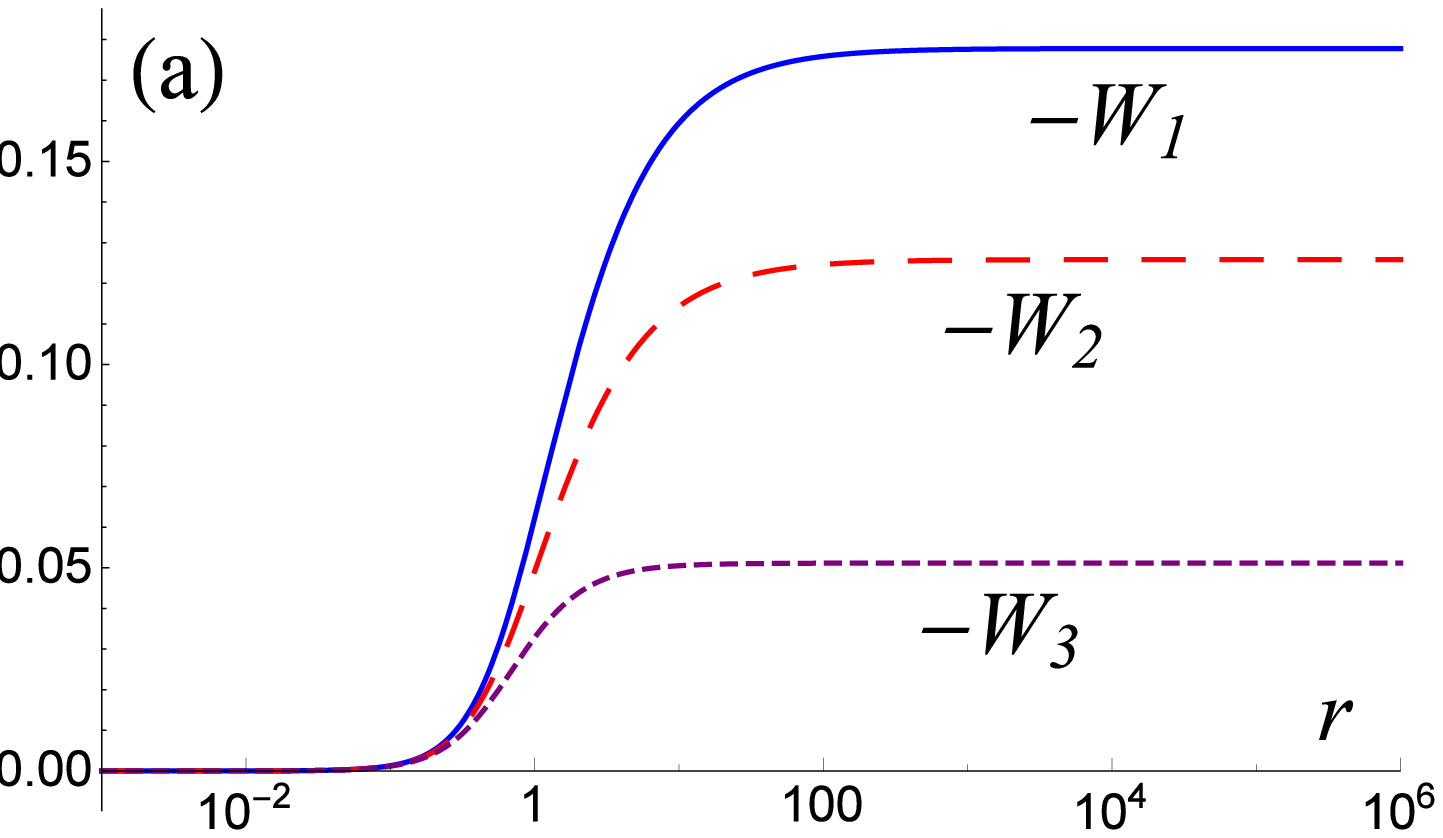}
\includegraphics[width=7.5cm]{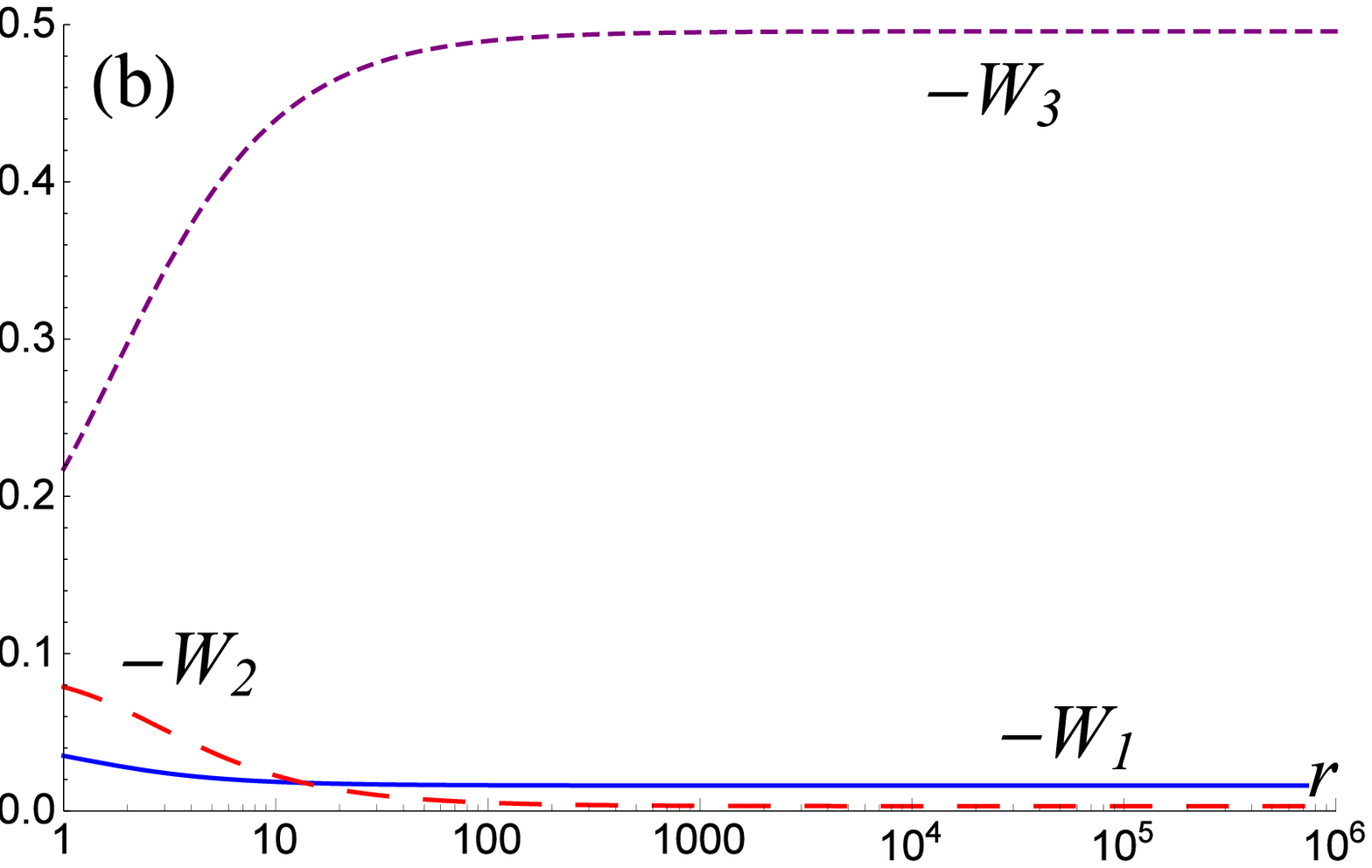}
\caption{$-W_{1}(r)$, $-W_{2}(r)$ and $-W_{3}(r)$ for the example nodeless  ${\mathfrak {su}}(4)$ solutions shown in figure \ref{fig:2}, (a) soliton and (b) black hole.
In both cases $-W_{j}\ge 0$ for all $r$, so that the inequalities (\ref{eq:sstabineqs}) are satisfied and the solutions have no instabilities in the
sphaleronic sector.}
\label{fig:4}
\end{figure}

By way of illustration, in figures \ref{fig:3} and \ref{fig:4}, we show how $-W_{j}$ (\ref{eq:Wdef}) depend on $r$ for the example soliton and black hole
solutions plotted in figures \ref{fig:1} and \ref{fig:2} respectively.
In all cases we see that $-W_{j}\ge 0$ for all $r$, so that these example solutions have no instabilities in the sphaleronic sector.

\section{Gravitational sector perturbations}
\label{sec:gravitational}

The gravitational sector consists of the even-parity Yang-Mills perturbations
$\delta \omega _{j}$ ($j=1,\ldots, N-1$) and the metric perturbations $\delta \mu $ and $\delta \Delta $.
The governing equations are the perturbed Einstein equations (\ref{eq:deltamup}--\ref{eq:deltadeltap}) together with the linearized Yang-Mills equations formed by subtracting equations (\ref{eq:qrealYMpert}) and
(\ref{eq:qimYMpert}), namely:
\begin{eqnarray}
0 & = &
-\delta {\ddot {\omega }}_j
+ {\ov {\mu }}^2 {\ov {S}}^2 \delta\omega''_j
+ {\ov {\mu }} {\ov {S}}^2
\left(
{\ov {\mu }}' + {\ov {\mu }} \frac{{\ov {S}}'}{{\ov {S}}}\right) \delta\omega_j'
+{\ov {\mu }} {\ov {S}}^2 {\ov {\omega }}_j'
\left[
\delta \mu' + {\ov {\mu }} \delta \left( \frac{S'}{S} \right)
+\delta \mu \frac{{\ov {S}}'}{{\ov {S}}}\right]
+{\ov {\mu }} {\ov {S}}^{2}{\ov {\omega }}_j''\delta\mu
\nonumber \\ & &
+\frac{{\ov {\mu }} {\ov {S}}^2}{r^2}
\left[
W_j \delta\omega_j -2{\ov {\omega }}_j^{2}\delta\omega_j
+{\ov {\omega }}_{j+1} {\ov {\omega }}_j \delta\omega_{j+1}
+{\ov {\omega }}_{j-1} {\ov {\omega }}_j \delta\omega_{j-1}
\right] .
\label{eq:deltaomegaeqn1}
\end{eqnarray}

\subsection{Metric perturbations}
\label{sec:metricperts}

The linearized Einstein equations (\ref{eq:deltamup}--\ref{eq:deltadeltap}) can be
used to eliminate the metric perturbations from the remaining gravitational sector
perturbation equations (\ref{eq:deltaomegaeqn1}). The linearized Einstein equation (\ref{eq:deltamudot}) can be immediately integrated to give
\begin{equation}
\delta \mu = -\frac {4{\ov {\mu }}}{r}
\sum _{j=1}^{N-1} {\ov {\omega }}_{j}' \, \delta \omega _{j}
+ \delta Y (r),
\end{equation}
where $\delta Y(r)$ is an arbitrary function of $r$ alone.
Using the linearized Einstein equation (\ref{eq:deltamup}) gives, after some lengthy algebra
\begin{equation}
\delta Y' = -\frac {1}{r} \left( 1 + 2{\ov {\Gamma }} \right) \delta Y ,
\label{eq:Yprime}
\end{equation}
where ${\ov {\Gamma }}$ is given by (\ref{eq:Pistatic}).
Integrating (\ref{eq:Yprime}) we find
\begin{equation}
\delta Y(r) = Y_{0} \exp \left( - \int _{r_{0}}^{r} \frac {1}{r'} \left[ 1+
2{\ov {\Gamma }}(r') \right] dr' \right)
\end{equation}
where $Y_{0}$ is a constant. The lower limit on the integral $r_{0}=0$ if we are considering perturbations of a static soliton solution, $r_{0}=r_{h}$ if we are considering perturbations of a static black hole solution.
Since we require our perturbations to vanish at either the origin or black hole event horizon, as relevant, it must be the case that $Y_{0}=0$ and hence
$\delta Y(r) \equiv 0$.
We therefore have
\begin{equation}
\delta \mu = -\frac {4{\ov {\mu }}}{r}
\sum _{j=1}^{N-1} {\ov {\omega }}_{j}' \, \delta \omega _{j}.
\label{eq:deltamu}
\end{equation}

\subsection{Gravitational sector perturbation equations in matrix form}
\label{sec:gravmatrix}

Now that we have the form (\ref{eq:deltamu}) of the metric perturbation
$\delta \mu $, together with (\ref{eq:deltadeltap}) for the perturbation $\delta \Delta '$, we can eliminate the metric perturbations from the gravitational sector perturbation equations (\ref{eq:deltaomegaeqn1}).

First we consider the quantity
\begin{equation}
\delta \mu ' + {\ov {\mu }} \delta \left( \frac {S'}{S} \right) +
\frac {{\ov {S}}'}{{\ov {S}}} \delta \mu
= -\frac {1}{r} \left(  \delta \mu + 2r^{2} \delta \Pi  \right),
\label{eq:gravuseful}
\end{equation}
where $\delta \Pi $ is given in (\ref{eq:deltaPi}). The right-hand-side of
(\ref{eq:gravuseful}) depends only on the perturbations $\delta \omega _{j}$ and not on their derivatives.
Next we define a vector of perturbations as follows:
\begin{equation}
{\bmath {\delta \omega }} = \left(
\delta \omega _{1}, \ldots , \delta \omega _{N-1} \right) ^{T}.
\end{equation}
Changing the radial co-ordinate to the tortoise co-ordinate (\ref{eq:tortoise}),
the gravitational sector perturbation equations (\ref{eq:deltaomegaeqn1}) take the form
\begin{equation}
-{\bmath {\delta }} {\ddot {\bmath {\omega }}}
=-\drs ^{2} {\bmath {\delta \omega }} + {\mathcal {M}} {\bmath {\delta \omega }} .
\label{eq:gravmatrix}
\end{equation}
The $\left( N-1 \right) \times \left( N-1 \right)$ matrix ${\mathcal {M}}$
depends only on the static equilibrium solutions and does not contain any derivative operators.
To simplify the entries of ${\mathcal {M}}$, we make extensive use of the static equilibrium field equations (\ref{eq:Ee}, \ref{eq:YMe}).
After a lengthy calculation, we can write the entries of the symmetric matrix ${\mathcal {M}}$ as follows.  There are three different types of entry which have different forms: (i) the diagonal entries ${\mathcal {M}}_{j,j}$, (ii) entries immediately above and below the diagonal ${\mathcal {M}}_{j,j+1}$ and (iii) other entries not on the diagonal nor immediately above or below it ${\mathcal {M}}_{j,k}$ ($k\neq j, j+1$). We give these entries explicitly below, where there is no summation:
\begin{eqnarray}
{\mathcal {M}}_{j,j} & = &
-\frac {{\ov {\mu }}{\ov {S}}^{2}}{r^{2}} \left[
W_{j} - 2{\ov {\omega }}_{j}^{2} \right]
- \frac {4{\mathcal {Q}}}{{\ov {\mu }}{\ov {S}}r }
\left( \drs {\ov {\omega }}_{j} \right) ^{2}
- \frac {8{\ov {S}}}{r^{3}}
W_{j} {\ov {\omega }}_{j} \drs {\ov {\omega }}_{j},
\nonumber
\\
{\mathcal {M}}_{j, j+1} & = &
-\frac {{\ov {\mu }}{\ov {S}}^{2}}{r^{2}}{\ov {\omega }}_{j} {\ov {\omega }}_{j+1}
-\frac {4{\mathcal {Q}}}{{\ov {\mu }}{\ov {S}}r }
\left( \drs {\ov {\omega }}_{j} \right) \left( \drs {\ov {\omega }}_{j+1} \right)
- \frac {4{\ov {S}}}{r^{3}}
\left[ W_{j} {\ov {\omega }}_{j} \drs {\ov {\omega }}_{j+1}
+ W_{j+1} {\ov {\omega }}_{j+1} \drs {\ov {\omega }}_{j} \right] ,
\nonumber
\\
{\mathcal {M}}_{j, k} & = &
-\frac {4{\mathcal {Q}}}{{\ov {\mu }}{\ov {S}}r }
\left(  \drs {\ov {\omega }}_{j} \right) \left( \drs {\ov {\omega }}_{k} \right)
-  \frac {4{\ov {S}}}{r^{3}}
\left[ W_{j} {\ov {\omega }}_{j} \drs {\ov {\omega }}_{k}
+ W_{k} {\ov {\omega }}_{k} \drs {\ov {\omega }}_{j} \right] ,
\label{eq:Mentries}
\end{eqnarray}
where we have defined
\begin{equation}
{\mathcal {Q}} =
\frac {1}{{\ov {\mu }}} \drs {\ov {\mu }} + \frac {1}{{\ov {S}}} \drs {\ov {S}}
+ \frac {{\ov {\mu }}{\ov {S}}}{r}.
\end{equation}

We now consider time-periodic perturbations for which ${\bmath {\delta \omega }}(t,r)=e^{i\sigma t} {\bmath {\delta \omega }}(r)$, and then the gravitational
sector perturbation equations (\ref{eq:gravmatrix}) are:
\begin{equation}
\sigma ^{2} {\bmath {\delta \omega }}
=-\drs ^{2} {\bmath {\delta \omega }} + {\mathcal {M}} {\bmath {\delta \omega }} .
\label{eq:gravtimepert}
\end{equation}
Since (\ref{eq:gravtimepert}) has the form of a standard Schr\"odinger-like equation, the operator on the right-hand-side of (\ref{eq:gravtimepert})
is positive if the matrix ${\mathcal {M}}$ is positive.
If this is the case, then $\sigma ^{2}$ is real and there are no unstable modes in the gravitational sector.

\subsection{Special cases}
\label{sec:Gspecial}

The gravitational sector perturbation equations (\ref{eq:gravmatrix}) are rather complicated in general, so first we consider some special cases.

\subsubsection{Schwarzschild-adS}
\label{sec:gSadS}

Setting ${\ov {\omega }}_{j} \equiv {\sqrt {j\left( N -j \right)}}$ for
$j=1,\ldots ,N-1$, the entries in the matrix ${\mathcal {M}}$ in the gravitational sector perturbation equations reduce to
\begin{eqnarray}
{\mathcal {M}}_{j,j} & = &
\frac {2{\ov {\mu }}{\ov {S}}^{2}}{r^{2}} j\left( N - j \right) ,
\nonumber \\
{\mathcal {M}}_{j,j+1} & = &
-\frac {{\ov {\mu }}{\ov {S}}^{2}}{r^{2}}
{\sqrt {j\left( N-j \right) \left( j+1 \right) \left( N -j - 1\right) }} ,
\nonumber \\
{\mathcal {M}}_{j,k} & = & 0, \qquad k \neq j, j+1.
\end{eqnarray}
Therefore we have
\begin{equation}
{\mathcal {M}} = \frac {{\ov {\mu }}{\ov {S}}^{2}}{r^{2}}  {\mathcal {E}}_{N-1},
\end{equation}
where ${\mathcal {E}}_{N-1}$ is the constant matrix with entries
\begin{eqnarray}
{\mathcal {E}}_{N-1,j,k} & = & {\sqrt {j\left( N - j\right)}}
{\sqrt {k\left( N-k\right) }}
\left[ 2\delta _{j,k} - \delta _{j+1,k} - \delta _{j-1,k} \right] .
\label{eq:MN-1def}
\end{eqnarray}
It is shown in Ref.~\onlinecite{Baxter2} that the eigenvalues of the matrix
${\mathcal {E}}_{N-1}$ are $k\left( k+1\right)$ for $k=1,\ldots N-1$,
so that the matrix ${\mathcal {E}}_{N-1}$ (and therefore the matrix ${\mathcal {M}}$)
is positive.
The net result of this is that, as anticipated, the embedded Schwarzschild-adS solution has no instabilities in the gravitational sector.

\subsubsection{Reissner-Nordstr\"om-adS}
\label{sec:gRNadS}

In this case ${\ov {\omega }}_{j} \equiv 0$ for all $j=1,\ldots , N-1$,
and the matrix ${\mathcal {M}}$ reduces to
\begin{equation}
{\mathcal {M}} = -\frac {{\ov {\mu }}{\ov {S}}^{2}}{r^{2}} {\mathcal {I}}_{N-1}.
\end{equation}
Therefore the matrix ${\mathcal {M}}$ is negative definite everywhere and, as expected, the embedded
magnetically charged Reissner-Nordstr\"om-adS solution is unstable.

\subsubsection{Embedded ${\mathfrak {su}}(2)$ solutions}
\label{sec:gsu2}

We now have ${\ov {\omega }}_{j} \equiv {\ov {\omega }}(r) {\sqrt {j\left( N-j \right) }}$
for $j=1,\ldots ,N-1$, and the entries of the matrix ${\mathcal {M}}$ (\ref{eq:Mentries}) take the form
\begin{widetext}
\begin{eqnarray}
{\mathcal {M}}_{j,j} & = &
\frac {{\ov {\mu }}{\ov {S}}^{2}}{r^{2}} \left( {\ov {\omega }} ^{2} -1 \right)
+ j \left( N - j \right) \left[
\frac {2{\ov {\mu }}{\ov {S}}^{2}}{r^{2}}  {\ov {\omega }}^{2}
 - \frac {4{\mathcal {Q}}}{{\ov {\mu }}{\ov {S}}r } \left( \drs {\ov {\omega }} \right) ^{2}
- \frac {8{\ov {S}}}{r^{3}}
\left( 1 -{\ov {\omega }} ^{2} \right) {\ov {\omega }}  \drs {\ov {\omega }} \right],
\nonumber \\
{\mathcal {M}}_{j,j+1} & = &
- {\sqrt {j\left( N-j \right) \left( j+1 \right) \left( N-j-1 \right) }}
\left[
\frac {{\ov {\mu }}{\ov {S}}^{2}}{r^{2}} {\ov {\omega }} ^{2}
+\frac {4{\mathcal {Q}}}{{\ov {\mu }}{\ov {S}}r }
\left( \drs {\ov {\omega }}  \right) ^{2}
+ \frac {8{\ov {S}}}{r^{3}}
\left( 1 - {\ov {\omega }}^{2} \right)  {\ov {\omega }}  \drs {\ov {\omega }}  \right]
,
\nonumber \\
{\mathcal {M}}_{j,k} & = &
- {\sqrt {j\left( N-j \right) k \left( N-k \right) }}
\left[
\frac {4{\mathcal {Q}}}{{\ov {\mu }}{\ov {S}}r }
\left(  \drs {\ov {\omega }}  \right) ^{2}
+  \frac {8{\ov {S}}}{r^{3}}
\left( 1- {\ov {\omega }}^{2} \right) {\ov {\omega }}  \drs  {\ov {\omega }} \right].
\end{eqnarray}
\end{widetext}
In this case it is helpful to consider the matrix ${\mathcal {M}}$ as a sum of three parts:
\begin{equation}
{\mathcal {M}}={\mathcal {N}}_{1} + {\mathcal {N}}_{2} + {\mathcal {N}}_{3},
\end{equation}
where
\begin{eqnarray}
{\mathcal {N}}_{1} & = &
\frac {{\ov {\mu }}{\ov {S}}^{2}}{r^{2}} \left( {\ov {\omega }}^{2} -1 \right)
{\mathcal {I}}_{N-1},
\nonumber \\
{\mathcal {N}}_{2} & = &
\frac {{\ov {\mu }}{\ov {S}}^{2}}{r^{2}}
{\ov {\omega }}^{2} {\mathcal {E}}_{N-1},
\nonumber \\
{\mathcal {N}}_{3} & = &
\left[
-\frac {4{\mathcal {Q}}}{{\ov {\mu }}{\ov {S}}r }
\left(  \drs {\ov {\omega }}  \right) ^{2}
-  \frac {8{\ov {S}}}{r^{3}}
\left( 1- {\ov {\omega }} ^{2} \right) {\ov {\omega }}  \drs {\ov {\omega }}
\right] {\widetilde {\mathcal {E}}}_{N-1} ,
\nonumber \\
\end{eqnarray}
where the constant matrix ${\mathcal {E}}_{N-1}$ is given in (\ref{eq:MN-1def})
and the constant matrix ${\widetilde {{\mathcal {E}}}}_{N-1}$ has entries
\begin{equation}
{\widetilde {{\mathcal {E}}}}_{N-1,j,k}=
{\sqrt {j\left( N-j \right) k \left( N-k \right) }}.
\end{equation}
The first matrix, ${\mathcal {N}}_{1}$, is positive if ${\ov {\omega }}(r) ^{2} \ge 1$ for all $r$, which is the same sufficient
condition as we found previously for no instabilities in the sphaleronic sector (see section \ref{sec:ssu2}).
Since the matrix ${\mathcal {E}}_{N-1}$ is positive, the second matrix
${\mathcal {N}}_{2}$ is also positive.
The positivity of the third matrix, ${\mathcal {N}}_{3}$ is less clear-cut.
However, it has been shown \cite{Winstanley1} that
${\sqrt {\left| \Lambda \right| }}{\ov {\omega }}' \rightarrow 0$ for all $r$ as $\left| \Lambda \right| \rightarrow \infty $.
Therefore, for sufficiently large $\left| \Lambda \right| $, the third matrix ${\mathcal {N}}_{3}$ is negligible compared with ${\mathcal {N}}_{1}$
and ${\mathcal {N}}_{2}$.
Hence, for sufficiently large $\left| \Lambda \right| $, if ${\ov {\omega }}(r)^{2}\ge 1$ for all $r$, the matrix ${\mathcal {M}}$ is positive and there
are no instabilities for embedded ${\mathfrak {su}}(2)$ solutions in the gravitational sector of ${\mathfrak {su}}(N)$ EYM perturbations.

In Ref.~\onlinecite{Winstanley1} it shown that ${\ov {\omega }}(r)^{2}\ge 1/3$ is a sufficient condition for ${\mathfrak {su}}(2)$ black holes to have no instabilities
in the gravitational sector of ${\mathfrak {su}}(2)$ EYM perturbations provided $\left| \Lambda \right| $ is sufficiently large.
As with the sphaleronic sector perturbations, for embedded ${\mathfrak {su}}(2)$ solutions in ${\mathfrak {su}}(N)$ EYM, we have a stronger
sufficient condition for the absence of unstable modes.
This is to be expected because of the greater number of degrees of freedom in the ${\mathfrak {su}}(N)$ gravitational sector perturbations than in the
${\mathfrak {su}}(2)$ gravitational sector perturbations.
We emphasize that the condition of ${\ov {\omega }}(r)^{2}\ge 1$ for all $r$ and sufficiently large $\left| \Lambda \right| $ is a sufficient condition, and there may be solutions which do not satisfy this condition but which are nonetheless stable.

\subsection{An alternative form of the gravitational sector perturbation equations}
\label{sec:gravalt}

In order to show, in the next section, that there exist both soliton and black hole non-embedded solutions of the static equilibrium field equations which have no instabilities in the gravitational sector, we shall follow the method of Ref.~\onlinecite{Sarbach2} and employ a nodal theorem due to Amann and Quittner \cite{Amann} which allows one to count the number of bound states of a Schr\"odinger-like equation.
In this subsection we state Amann and Quittner's result and cast our gravitational sector perturbation equations (\ref{eq:gravtimepert}) in the form required for the application of the theorem in section \ref{sec:gravstab}. We will need to consider solitons and black holes separately.

\subsubsection{The nodal theorem}
\label{sec:nodal}

Amann and Quittner's theorem \cite{Amann} is concerned with the number of bound states of a radial Schr\"odinger-like operator.
Let ${\mathfrak {D}}$ be the linear differential operator
\begin{equation}
{\mathfrak {D}} {\bmath  {u}} = -\frac {d}{d\rho } \left[ {\mathfrak {A}} (\rho )
\frac {d}{d\rho } {\bmath {u}} \right]
+ \left[ \frac {1}{\rho ^{2}} {\mathfrak {B}}(\rho ) + {\mathfrak {C}} (\rho )
\right] {\bmath {u}},
\label{eq:operator}
\end{equation}
acting on $n$-dimensional vectors ${\bmath {u}}(\rho )$, where $\rho \in \left[ 0,
\infty \right) $ lies on the half-line.
The $n\times n$ matrices ${\mathfrak {A}}(\rho )$, ${\mathfrak {B}}(\rho )$ and
${\mathfrak {C}}(\rho )$ are assumed to be real, symmetric, smooth and uniformly bounded on $\left[ 0, \infty \right) $.
It is further assumed that ${\mathfrak {A}}(\rho )$ is uniformly positive definite on $\left[ 0, \infty \right) $, that is,
there is a constant ${\mathfrak {a}}>0$ such that
\begin{equation}
{\mathfrak {A}}(\rho ) \ge {\mathfrak {a}} >0 \qquad
{\mbox {for $0\le \rho < \infty $}},
\label{eq:Acond}
\end{equation}
and that ${\mathfrak {B}}(0)$ is non-negative.

The theorem is concerned with the eigenvalue problem
\begin{equation}
{\mathfrak {D}} {\bmath {u}} = \lambda {\bmath {u}}, \qquad
{\bmath {u}}(0)=0,
\label{eq:eigenvalue}
\end{equation}
where ${\bmath {u}} \in L_{2}\left( (0,\infty ), {\mathbb {R}}^{n} \right) $.
Following Ref.~\onlinecite{Amann}, we further assume that the bottom of the essential spectrum of ${\mathfrak {D}}$ is positive and that the eigenvalue problem (\ref{eq:eigenvalue}) has only finitely many negative eigenvalues.
Sufficient conditions for this assumption to be valid are \cite{Amann}:
\begin{equation}
{\mathfrak {C}}(\rho ) \rightarrow 0 {\mbox { as $\rho \rightarrow \infty $}}
\label{eq:Ccond1}
\end{equation}
and
\begin{equation}
{\mathfrak {C}}(\rho ) \rho^{2} + {\mathfrak {B}}(\rho ) \ge - {\mathfrak {b}}
\label{eq:Ccond2}
\end{equation}
for some ${\mathfrak {b}}< 1/4$ and all sufficiently large $\rho $.

The statement of Amann and Quittner's theorem involves an auxiliary problem, which we now state.
Choose $n$ linearly independent real, constant, $n$-dimensional vectors
${\bmath {e}}_{j}$, $j=1,\ldots, n$.
Let ${\mathfrak {c}}>0$ and let
\begin{equation}
{\mathfrak {U}}_{{\mathfrak {c}}}= \left[ {\bmath {u}}_{1},  \ldots ,
{\bmath {u}}_{n} \right]
\end{equation}
be the $n\times n$ matrix whose columns are the solutions of the $n$ initial
value problems
\begin{equation}
{\mathfrak {D}} {\bmath {u}}_{j} = 0, \quad
{\mathfrak {c}}< \rho < \infty,
\quad
{\bmath {u}}_{j} ({\mathfrak {c}}) =0, \quad
\frac {d}{d\rho }{\bmath {u}}_{j}({\mathfrak {c}}) = {\bmath {e}}_{j},
\label{eq:IVP}
\end{equation}
for $j=1,\ldots ,n$.
We then define a scalar function ${\mathfrak {F}}(\rho )$ by
\begin{equation}
{\mathfrak {F}}(\rho )= \det {\mathfrak {U}}_{\mathfrak {c}}(\rho ).
\label{eq:Fdef}
\end{equation}

We are now in a position to quote Amann and Quittner's theorem \cite{Amann}:
\begin{theorem}
If ${\mathfrak {c}}>0$ is sufficiently small and ${\mathfrak {d}}>{\mathfrak {c}}$
is sufficiently large, the number of zeros (counted with multiplicities) in the interval $({\mathfrak {c}},{\mathfrak {d}})$ of the function ${\mathfrak {F}}(\rho )$
equals the number of negative eigenvalues of (\ref{eq:eigenvalue}) (counted with multiplicities).
\end{theorem}

In order to apply this theorem, we need to cast the gravitational sector perturbation equations (\ref{eq:gravtimepert}) in the form (\ref{eq:eigenvalue}), choose a suitable co-ordinate $\rho $ and check that the matrices ${\mathfrak {A}}(\rho )$,
${\mathfrak {B}}(\rho )$ and ${\mathfrak {C}}(\rho )$ satisfy the required conditions
(\ref{eq:Acond}, \ref{eq:Ccond1}, \ref{eq:Ccond2}) together with the requirement that ${\mathfrak {B}}(0)$ is non-negative.
To do this, as in Ref.~\onlinecite{Sarbach2}, we need to consider soliton and black hole solutions separately.  We consider black holes first as this case is simpler.

\subsubsection{Black holes}
\label{sec:nodalBHs}

Following Ref.~\onlinecite{Sarbach2}, for static black hole solutions we take
\begin{equation}
\rho = -r_{*} \in \left[ 0,\infty \right)
\end{equation}
where $r_{*}$ is the tortoise co-ordinate (\ref{eq:tortoise}).
This means that $\rho \rightarrow 0$ corresponds to $r\rightarrow \infty $,
and $\rho \rightarrow \infty $ corresponds to $r\rightarrow r_{h}$, approaching the event horizon.

We choose the $\left( N-1 \right) \times \left( N-1 \right) $ matrices appearing in the differential operator ${\mathfrak {D}}$
(\ref{eq:operator}) as follows:
\begin{equation}
{\mathfrak {A}}(\rho ) = {\mathcal {I}}_{N-1},
\qquad
{\mathfrak {B}}(\rho ) =0,
\qquad
{\mathfrak {C}}(\rho ) = {\mathcal {M}},
\end{equation}
where the matrix ${\mathcal {M}}$ has entries (\ref{eq:Mentries}).
Taking ${\mathfrak {a}}=\frac {1}{2}>0$, the condition (\ref{eq:Acond}) is automatically satisfied, and, furthermore, ${\mathfrak {B}}(0)=0$ is non-negative.
It remains therefore to check the conditions on the matrix ${\mathfrak {C}}$
(\ref{eq:Ccond1}--\ref{eq:Ccond2}).

We first examine the behaviour of the matrix ${\mathfrak {C}}(\rho )$ as
$\rho \rightarrow 0$, that is, $r\rightarrow \infty $.
Using the boundary conditions (\ref{eq:infinity}),
and noting that, as $r\rightarrow \infty $,
\begin{equation}
\drs {\ov {\omega }}_{j} = \frac {\Lambda c_{j}}{3} +O(r^{-1}),
\qquad
{\mathcal {Q}} = -\Lambda r + O(1),
\end{equation}
we find that the leading-order behaviour of ${\mathfrak {C}}$ is given by the entries (\ref{eq:Mentries})
\begin{eqnarray}
{\mathcal {M}}_{j,j} & = &
\frac {\Lambda }{3}\left[ 1- 3{\ov {\omega }}_{j,\infty }^{2}
+\frac {1}{2} \left( {\ov {\omega }}_{j+1,\infty }^{2}
+{\ov {\omega }}_{j-1,\infty }^{2} \right) \right]
+ O(r^{-1}),
\nonumber \\
{\mathcal {M}}_{j,j+1}  & = &
\frac {\Lambda }{3} {\ov {\omega }}_{j,\infty }{\ov {\omega }}_{j+1,\infty }
+ O(r^{-1}),
\nonumber \\
{\mathcal {M}}_{j,k} & = &
O(r^{-2}).
\end{eqnarray}
Therefore the matrix ${\mathfrak {C}}(\rho )$ remains bounded as
$r\rightarrow \infty $, that is, $\rho \rightarrow 0$.

As $\rho \rightarrow \infty $, we have $r\rightarrow r_{h}$.
Using the boundary conditions (\ref{eq:horizon}), we find that
\begin{equation}
\drs {\ov {\omega }}_{j} = O(r-r_{h}), \qquad
{\mathcal {Q}} = {\ov {S}}(r_{h}) {\ov {\mu }}'(r_{h}) + O(r-r_{h}),
\end{equation}
and that the entries of the matrix ${\mathfrak {C}}$ are all $O(r-r_{h})$ as
$r\rightarrow r_{h}$, so that ${\mathfrak {C}}(\rho ) \rightarrow 0$ as $\rho \rightarrow \infty $, satisfying (\ref{eq:Ccond1}).
This also means that the matrix ${\mathfrak {C}}(\rho )$ is uniformly bounded
on $\left[ 0, \infty \right) $.

To check whether (\ref{eq:Ccond2}) is satisfied, we first note that, for $r\sim r_{h}$,
\begin{equation}
\rho = -r_{*} \sim -\rho _{h}\ln \left( r- r_{h} \right) ,
\end{equation}
where $\rho _{h}$ is a positive constant.
Therefore, as $\rho \rightarrow \infty $,
\begin{equation}
{\mathfrak {C}}(\rho ) \sim O\left( e^{-\frac {\rho }{\rho _{h}}} \right) .
\end{equation}
Therefore $\rho ^{2} {\mathfrak {C}}(\rho ) \rightarrow 0$ as
$\rho \rightarrow \infty $.
Therefore there exists a $\rho _{1}$ such that for all $\rho > \rho _{1}$,
\begin{equation}
\frac {1}{8} < \rho ^{2} {\mathfrak {C}} (\rho ) < -\frac {1}{8} .
\end{equation}
Hence we have satisfied (\ref{eq:Ccond2}) with ${\mathfrak {b}}=1/8$.

Therefore we have cast the gravitational sector perturbation equations in the form (\ref{eq:eigenvalue}) required for the application of the nodal theorem, and all the conditions required by the theorem are satisfied.
We comment that this case was simpler to deal with than the situation in Ref.~\onlinecite{Sarbach2}, because in that paper terms arising from non-spherically symmetric perturbations cannot be included in ${\mathfrak {C}}(\rho )$.

\subsubsection{Solitons}
\label{sec:nodalsolitons}

For solitons, the system of gravitational perturbation equations needs further transformation before it is in the form required for the application of the nodal theorem.

Following Ref.~\onlinecite{Sarbach2}, we define
\begin{equation}
\rho = r^{-\frac {1}{2}},
\end{equation}
so that $\rho \rightarrow 0$ corresponds to $r\rightarrow \infty $ and
$\rho \rightarrow \infty $ corresponds to the origin.
We also make a transformation of the perturbations:
\begin{equation}
{\bmath {\delta \omega }} = {\mathfrak {X}}{\bmath {v}},
\end{equation}
where
\begin{equation}
{\mathfrak {X}} = r^{\frac {3}{4}} \left(
{\ov {\mu }}{\ov {S}} \right) ^{-\frac {1}{2}} .
\end{equation}
The gravitational sector perturbation equations now take the form
\begin{equation}
4{\mathfrak {X}}^{4} \lambda {\bmath {v}}
= -\frac {d^{2}{\bmath {v}}}{d\rho ^{2}} + {\widetilde {\mathcal {M}}}{\bmath {v}},
\label{eq:gravmodified}
\end{equation}
where
\begin{equation}
{\widetilde {{\mathcal {M}}}}= 4{\mathfrak {X}}^{4}
{\mathcal {M}}
- {\mathfrak {X}} \frac {d}{d\rho } \left( \frac {1}{{\mathfrak {X}}^{2}}
\frac {d{\mathfrak {X}}}{d\rho } \right) {\mathcal {I}}_{N-1}.
\end{equation}
Comparing with (\ref{eq:operator}), as with the black hole case we take
\begin{equation}
{\mathfrak {A}}(\rho ) ={\mathcal {I}}_{N-1}.
\end{equation}
To fix the matrices ${\mathfrak {B}}(\rho )$ and ${\mathfrak {C}}(\rho )$, we
need to study the behaviour of the matrix ${\widetilde {\mathcal {M}}}$ as
$\rho \rightarrow 0$.

As $\rho \rightarrow 0$, $r\rightarrow \infty $ and
\begin{equation}
{\mathfrak {X}} = \frac {{\sqrt {3}}}{{\sqrt {-\Lambda }}}r^{-\frac {1}{4}}
+O\left( r^{-\frac {5}{4}} \right)
= \frac {{\sqrt {3}}}{{\sqrt {-\Lambda }}}\rho ^{\frac {1}{2}}
+O\left( \rho ^{\frac {5}{2}} \right) .
\label{eq:xrho0}
\end{equation}
Therefore, as $\rho \rightarrow 0$,
\begin{equation}
{\mathfrak {X}} \frac {d}{d\rho } \left( \frac {1}{{\mathfrak {X}}^{2}}
\frac {d{\mathfrak {X}}}{d\rho } \right)
= -\frac {3}{4} \rho ^{-2} + O(1).
\end{equation}
From the analysis of the previous subsection,
we know that ${\mathcal {M}}=O(1)$ as $r\rightarrow \infty $.
This suggests that we should take
\begin{equation}
{\mathfrak {B}}(\rho ) = -\rho ^{2} {\mathfrak {X}} \frac {d}{d\rho } \left( \frac {1}{{\mathfrak {X}}^{2}}
\frac {d{\mathfrak {X}}}{d\rho } \right) {\mathcal {I}}_{N-1},
\qquad
{\mathfrak {C}}(\rho ) = 4{\mathfrak {X}}^{4}
{\mathcal {M}} .
\label{eq:BCsoliton}
\end{equation}
With this choice, we have
\begin{equation}
{\mathfrak {B}}(0) = \frac {3}{4}{\mathcal {I}}_{N-1},
\end{equation}
which is non-negative as required.

To check the other conditions (\ref{eq:Ccond1}, \ref{eq:Ccond2}) on the matrices ${\mathfrak {B}}(\rho )$ and
${\mathfrak {C}}(\rho )$, we need to examine their behaviour as $\rho \rightarrow \infty $, that is, $r\rightarrow 0$.
In this case, using the boundary conditions (\ref{eq:origin1})
\begin{equation}
{\mathfrak {X}} = \frac {1}{{\sqrt {S_{0}}}}
r^{\frac {3}{4}}+O\left( r^{\frac {7}{4}} \right)
= \frac {1}{{\sqrt {S_{0}}}}
\rho ^{-\frac {3}{2}}+O\left( \rho ^{-\frac {7}{2}} \right) ,
\label{eq:Xrhoinf}
\end{equation}
which gives
\begin{equation}
{\mathfrak {X}} \frac {d}{d\rho } \left( \frac {1}{{\mathfrak {X}}^{2}}
\frac {d{\mathfrak {X}}}{d\rho } \right)
= -\frac {3}{4} \rho ^{-2} + O(\rho ^{-4}).
\label{eq:Xdrhoinf}
\end{equation}
Therefore ${\mathfrak {B}}(\rho ) \rightarrow \frac {3}{4}{\mathcal {I}}_{N-1}$ as
$\rho \rightarrow \infty $
and the matrix ${\mathfrak {B}}(\rho )$ is uniformly bounded on
$\left[ 0, \infty  \right) $.

We next turn to the behaviour of the matrix ${\mathfrak {C}}(\rho )$ as $\rho \rightarrow 0$ and $r\rightarrow \infty $.
We have already seen that ${\mathcal {M}} = O(1)$ as $r\rightarrow \infty $.
Then, using the definition of ${\mathfrak {C}}(\rho )$ (\ref{eq:BCsoliton}) and the behaviour of ${\mathfrak {X}}$ as $\rho \rightarrow 0$ (\ref{eq:xrho0}),
the matrix ${\mathfrak {C}}(\rho ) \rightarrow 0$ as $\rho \rightarrow 0$.

Therefore it remains to investigate the properties of ${\mathfrak {C}}(\rho )$
as $\rho \rightarrow \infty $ and $r\rightarrow 0$.
Using the boundary conditions (\ref{eq:origin1}),
we first note that, as $r\rightarrow 0$,
\begin{equation}
\drs {\ov {\omega }}_{j} = O(r), \qquad
W_{j} =O(r^{2}), \qquad
{\mathcal {Q}} = \frac {S_{0}}{r} + O(1).
\end{equation}
Therefore the behaviour of the entries of the matrix ${\mathcal {M}}$ (\ref{eq:Mentries}) as $r\rightarrow 0$ is:
\begin{eqnarray}
{\mathcal {M}}_{j,j} & = &
\frac {2S_{0}^{2}}{r^{2}} j\left( N-j\right) +O(1),
\nonumber \\
{\mathcal {M}}_{j,j+1} & = & - \frac {S_{0}^{2}}{r^{2}}
{\sqrt {j\left( N- j\right) \left( j +1 \right) \left( N-j-1 \right) }}
+O(1),
\nonumber \\
{\mathcal {M}}_{j,k} & = & O(1).
\label{eq:Mrhoinf}
\end{eqnarray}
Using the behaviour of ${\mathfrak {X}}$ as $\rho \rightarrow \infty $ (\ref{eq:Xrhoinf}), we then have
\begin{equation}
{\mathfrak {C}} (\rho ) = O(\rho ^{-2}) \rightarrow 0 {\mbox { as $\rho \rightarrow \infty $}},
\end{equation}
so (\ref{eq:Ccond1}) is satisfied.

Looking at the remaining condition (\ref{eq:Ccond2}), using the asymptotic forms
(\ref{eq:Xrhoinf}, \ref{eq:Xdrhoinf}, \ref{eq:Mrhoinf}), we see that, as $\rho \rightarrow \infty $,
\begin{equation}
{\mathfrak {B}}(\rho ) + \rho ^{2} {\mathfrak {C}}(\rho )
= \frac {3}{4} {\mathcal {I}}_{N-1}
+4{\mathcal {E}}_{N-1}
+O(\rho ^{-2})
\end{equation}
where the matrix ${\mathcal {E}}_{N-1}$ is given by (\ref{eq:MN-1def}).
Since we know that the matrix ${\mathcal {E}}_{N-1}$ has only positive eigenvalues,
we deduce that ${\mathfrak {B}}(\rho )+\rho ^{2}{\mathfrak {C}}(\rho)$ is positive for sufficiently large $\rho  $ and therefore (\ref{eq:Ccond2}) is satisfied.

Finally in this subsection we note that the eigenvalue problem we have in the gravitational sector (\ref{eq:gravmodified}) is not exactly of the form
(\ref{eq:eigenvalue}) required for the application of the nodal theorem.
However, since ${\mathfrak {X}}^{4} \ge 0$ everywhere, this will not be a major difficulty in our analysis in section \ref{sec:gravstab}.

\subsection{Existence of static solutions with no gravitational sector instabilities}
\label{sec:gravstab}

We are now in a position to prove, in this subsection, the existence of non-trivial (that is, non-embedded) ${\mathfrak {su}}(N)$ solitons and black holes which have no instabilities in the gravitational sector.
For both solitons and black holes, our argument will use Amann and Quittner's nodal theorem \cite{Amann}.

For black holes, in section \ref{sec:nodalBHs} we have written the gravitational sector perturbation equations in the standard form (\ref{eq:eigenvalue})
required for the application of the nodal theorem.
In order to show that black hole solutions have no instabilities in the gravitational sector, it therefore suffices to show that the function
${\mathfrak {F}}(\rho )$ (\ref{eq:Fdef}) has no zeros on an interval $\rho \in ({\mathfrak {c}},{\mathfrak {d}})$, for small
${\mathfrak {c}}$ and large ${\mathfrak {d}}$.

For soliton solutions, the argument is a little more involved.
First of all, the gravitational sector perturbation equations (\ref{eq:gravmodified}) take the form
\begin{equation}
{\mathfrak {G}} \lambda {\bmath {v}}
= {\mathfrak {D}}{\bmath {v}}
= -\frac {d^{2}{\bmath {v}}}{d\rho ^{2}} + {\widetilde {\mathcal {M}}}{\bmath {v}},
\label{eq:eigenvaluemod}
\end{equation}
where ${\mathfrak {G}}=4{\mathfrak {X}}^{4}$ is a positive function, whereas the nodal theorem applies to the eigenvalue problem
${\mathfrak {D}}{\bmath {v}}=\lambda {\bmath {v}}$ (\ref{eq:eigenvalue}).
Suppose that we are able to show that there exist ${\mathfrak {su}}(N)$ soliton solutions for which the function ${\mathfrak {F}}(\rho )$ (\ref{eq:Fdef}) has
no zeros in the interval $\rho \in ({\mathfrak {c}},{\mathfrak {d}})$.
Then, applying the nodal theorem, the eigenvalue problem (\ref{eq:eigenvalue}) has no negative eigenvalues.
This means that the operator ${\mathfrak {D}}$ is a positive operator.
Then, if ${\mathfrak {D}}$ is a positive operator, it must be the case that the eigenvalue problem (\ref{eq:eigenvaluemod}) also cannot have
any negative eigenvalues because ${\mathfrak {G}}$ is a positive function.
The upshot is that, for the soliton case as for the black hole case, if we can show that the function ${\mathfrak {F}}(\rho )$ (\ref{eq:Fdef})
has no zeros in
an appropriate interval, then there are no instabilities in the gravitational sector.

From the existence theorems in  Ref.~\onlinecite{Baxter3}, we know that the equilibrium ${\mathfrak {su}}(N)$ solutions of the field equations are analytic in $r$, $\Lambda $ and the parameters at the origin or event horizon characterizing either soliton or black hole solutions.
The matrices ${\mathfrak {A}}$, ${\mathfrak {B}}$ and ${\mathfrak {C}}$ appearing in the operator ${\mathfrak {D}}$ (\ref{eq:operator})
are analytic functions of the equilibrium field functions ${\ov {\mu }}(r)$, ${\ov {S}}(r)$ and ${\ov {\omega }}_{j}(r)$ and $r$ (and hence $\rho $)
for values of $\rho $ in our interval of interest $({\mathfrak {c}}, {\mathfrak {d}})$.
Standard existence theorems for ordinary differential equations (see, for example, Ref.~\onlinecite{codlev}) then tell us that the solutions ${\bmath {u}}_{j}$
of the initial value problems (\ref{eq:IVP}) are also analytic functions of $\rho $, $\Lambda $ and the initial parameters at either the origin or
event horizon. Therefore, the function ${\mathfrak {F}}(\rho )$ (\ref{eq:Fdef}) is also analytic in $\rho $, $\Lambda $ and the initial parameters at the origin or event horizon.

In section \ref{sec:ssu2}, we proved the existence of embedded ${\mathfrak {su}}(2)$ solitons and black holes
for which ${\ov {\omega }}(r)^{2}>1$ for all $r$.
There we also showed that these embedded ${\mathfrak {su}}(2)$ solutions have no instabilities in the sphaleronic sector.
From section \ref{sec:gsu2} they also have no instabilities in the gravitational sector.
Pick such an embedded ${\mathfrak {su}}(2)$ solution (either a soliton or a black hole).
Fix ${\mathfrak {c}}$ to be very small and ${\mathfrak {d}}$ to be very large.
Then, from the nodal theorem, the function ${\mathfrak {F}}$ will have no zeros on the interval $({\mathfrak {c}},{\mathfrak {d}})$ for this particular embedded ${\mathfrak {su}}(2)$ soliton or black hole.
From the existence theorems in Ref.~\onlinecite{Baxter3}, there exist genuinely ${\mathfrak {su}}(N)$ solutions in a neighbourhood of this embedded
${\mathfrak {su}}(2)$ solution.
Since ${\mathfrak {F}}(\rho )$ is analytic in the parameters at the origin or event horizon which characterize the ${\mathfrak {su}}(N)$ solutions,
providing the ${\mathfrak {su}}(N)$ solutions are sufficiently close to the embedded ${\mathfrak {su}}(2)$ solution, the function ${\mathfrak {F}}(\rho )$
will continue to have no zeros in the interval $({\mathfrak {c}},{\mathfrak {d}})$ for the ${\mathfrak {su}}(N)$ solitons or black holes.

Therefore, if we consider ${\mathfrak {su}}(N)$ solutions sufficiently close to this stable embedded ${\mathfrak {su}}(2)$ solution,
using the nodal theorem (and
considering the operator ${\mathfrak {D}}$ for soliton solutions as described above), we have therefore proven that these ${\mathfrak {su}}(N)$ solutions
have no instabilities in the gravitational sector.

In section \ref{sec:sphstab}, we showed that ${\mathfrak {su}}(N)$ solutions in a neighbourhood ${\mathfrak {N}}_{1}$ of the above stable embedded ${\mathfrak {su}}(2)$ solution
have no instabilities in the sphaleronic sector.
Having, in the current section, shown that ${\mathfrak {su}}(N)$ solutions in another neighbourhood ${\mathfrak {N}}_{2}$ of the above
embedded ${\mathfrak {su}}(2)$ solution
have no instabilities in the gravitational sector, we can deduce that those ${\mathfrak {su}}(N)$ solutions in the intersection of ${\mathfrak {N}}_{1}$
and ${\mathfrak {N}}_{2}$ are stable under linear, spherically symmetric perturbations.

\section{Conclusions}
\label{sec:conc}

In this paper we have proven the existence of non-trivial, purely magnetic, spherically symmetric,
${\mathfrak {su}}(N)$ Einstein-Yang-Mills solitons and black holes in asymptotically
anti-de Sitter space (with a negative cosmological constant $\Lambda $) which are stable under linear, spherically symmetric perturbations.

The equilibrium solutions we consider are purely magnetic and spherically symmetric and the Yang-Mills field is described by $N-1$ functions
${\ov {\omega }}_{j}(r)$.
With an appropriate choice of gauge, the perturbation equations for linear, spherically symmetric, perturbations decouple into two sectors: the sphaleronic
sector and the gravitational sector.
The sphaleronic sector, involving only gauge field perturbations, is easier to analyze and is considered in section \ref{sec:sphaleronic}.
We find a series of inequalities (\ref{eq:sstabineqs}) on the equilibrium functions ${\ov {\omega }}_{j}(r)$ which are sufficient for there to be no instabilities in the sphaleronic sector.
We first proved the existence of embedded ${\mathfrak {su}}(2)$ solutions which satisfy these inequalities, before showing that ${\mathfrak {su}}(N)$
solutions in a neighbourhood of these stable embedded ${\mathfrak {su}}(2)$ solutions also have no instabilities in the sphaleronic sector.

The gravitational sector is studied in section \ref{sec:gravitational}.
The metric perturbations can be eliminated to leave a set of equations for gauge field perturbations.
Our approach to proving stability in this sector follows Ref.~\onlinecite{Sarbach2}, making use of a nodal theorem for a multidimensional Schr\"odinger system \cite{Amann}.
Again we can prove the existence of ${\mathfrak {su}}(N)$ solutions, in a neighbourhood of stable embedded ${\mathfrak {su}}(2)$ solutions, which have no instabilities in the gravitational sector.

A natural question is how the stable ${\mathfrak {su}}(N)$ EYM black holes whose existence we have proven in this paper fit into the context of the ``no-hair'' conjecture as formulated by Bizon \cite{Bizon2}:
\begin{quotation}
Within a given matter model, a {\em {stable}} stationary black hole is uniquely
determined by global charges.
\end{quotation}
It is argued in Ref.~\onlinecite{Shepherd} that, for sufficiently large $\left| \Lambda \right| $, there exist $N-1$ non-Abelian magnetic global charges which
uniquely characterize ${\mathfrak {su}}(N)$ EYM black holes, at least for large event horizon radius $r_{h}$ and in a region of the parameter space which
contains embedded ${\mathfrak {su}}(2)$ black holes.
For both the sphaleronic and gravitational sectors, our proof of the existence of stable ${\mathfrak {su}}(N)$ EYM black holes (and solitons)
is valid for large $\left| \Lambda \right| $.
Combining our results in this paper with those in Ref.~\onlinecite{Shepherd}, we have evidence that at least some large stable
${\mathfrak {su}}(N)$ EYM black holes are uniquely determined by global charges, in accordance with Bizon's ``no-hair'' conjecture
(see also Ref.~\onlinecite{Winstanley}).

In this paper we have considered only purely magnetic, spherically symmetric solitons and black holes.
The existence of ${\mathfrak {su}}(N)$ purely magnetic topological black holes has been proven \cite{Baxter4}, and
solutions found numerically for the ${\mathfrak {su}}(3)$ gauge group \cite{Baxter5}.
Very recently it has been shown that the argument we have presented here can be extended to show the stability of some of
these ${\mathfrak {su}}(N)$ purely magnetic topological black holes \cite{Baxter6}.
Dyonic solitons and black holes
in ${\mathfrak {su}}(2)$ EYM have been found numerically \cite{Bjoraker} and the existence of solutions where
both the electric and magnetic gauge field functions have no zeros has been proven \cite{Nolan}.
Dyonic solutions have also been found numerically for the larger gauge group ${\mathfrak {su}}(3)$ \cite{Shepherd1}.
Recently the existence of dyonic soliton and black hole solutions of the ${\mathfrak {su}}(N)$ field equations has been proven \cite{Baxter7}.
The existence of stable  ${\mathfrak {su}}(2)$ dyonic solutions has been proven very recently \cite{Nolan1} and it would be interesting to investigate
whether our results in this paper on the existence of stable purely magnetic solitons and black holes in ${\mathfrak {su}}(N)$ EYM in
anti-de Sitter space can be extended to dyonic solutions.

Finally, we comment that in this paper our focus has been the classical stability of ${\mathfrak {su}}(N)$ purely magnetic EYM black holes and solitons in anti-de Sitter space.
We have considered only linear, spherically symmetric perturbations.
The extension of our results to general linear perturbations is likely to be extremely challenging technically (see Refs.~\onlinecite{Sarbach1,Sarbach2} for the
${\mathfrak {su}}(2)$ case) and we would expect that at least some of the solutions which are stable under spherically symmetric linear perturbations will
remain stable when general linear perturbations are considered.
Going beyond classical stability, recent work has considered the thermodynamics of
purely magnetic ${\mathfrak {su}}(2)$ EYM black holes in anti-de Sitter space \cite{thermo} (see also Refs.~\onlinecite{adSthermo,Shepherd}).
In the ${\mathfrak {su}}(2)$ EYM case, for generic (non-integer) magnetic charge there are two branches of asymptotically anti-de Sitter black holes,
one of which is thermodynamically stable.
It would be interesting to extend the work of Ref.~\onlinecite{thermo} to the larger ${\mathfrak {su}}(N)$ gauge group.

\begin{acknowledgments}
We thank Brien Nolan for many helpful discussions on this project, and the anonymous referee for helpful comments.
The work of E.W. is supported by the Lancaster-Manchester-Sheffield
Consortium for Fundamental Physics under STFC grant ST/L000520/1.
\end{acknowledgments}


\end{document}